\documentclass[pdflatex,bst/sn-mathphys-num]{sn-jnl}% Math and Physical Sciences Numbered Reference Style
\usepackage[numbers]{natbib}
\usepackage{graphicx}%
\usepackage{multirow}%
\usepackage{amsmath,amssymb,amsfonts}%
\usepackage{amsthm}%
\usepackage{mathrsfs}%
\usepackage[title]{appendix}%
\usepackage{xcolor}%
\usepackage{textcomp}%
\usepackage{manyfoot}%
\usepackage{booktabs}%
\usepackage{algorithm}%
\usepackage{algorithmicx}%
\usepackage{algpseudocode}%
\usepackage{listings}%
\usepackage{subcaption} % 用于子图排版
\usepackage{tcolorbox}  % 用于画框
\tcbuselibrary{listings, skins} % 必须加载 listings 和 skins 库
\theoremstyle{thmstyleone}%
\theoremstyle{thmstyletwo}%

\theoremstyle{thmstylethree}%

\begin{document}

\title[Article Title]{From Literature to Lab: Closed-Loop Advancement of Perovskite Solar Cells via Domain Knowledge Guided LLM}


\author[1]{\fnm{Penglei Sun}}\email{psun012@connect.hkust-gz.edu.cn}
\equalcont{These authors contributed equally to this work.}

\author[2]{\fnm{Shuyan Chen}}\email{schen068@connect.hkust-gz.edu.cn}
\equalcont{These authors contributed equally to this work.}

\author[1]{\fnm{Xiang Liu}}\email{xliu886@connect.hkust-gz.edu.cn}

\author[2]{\fnm{Longhan Zhang}}\email{lzhang619@connect.hkust-gz.edu.cn}

\author[1]{\fnm{Huajie You}}\email{hyou603@connect.hkust-gz.edu.cn}

\author*[2,3]{\fnm{Chang Yan}}\email{changyan@hkust-gz.edu.cn}

\author[1]{\fnm{Yongqi Zhang}}\email{yongqizhang@hkust-gz.edu.cn}

\author*[1,3]{\fnm{Xiaowen Chu}}\email{xwchu@hkust-gz.edu.cn}

\author*[2,3,4]{\fnm{Tong-yi Zhang}}\email{mezhangt@hkust-gz.edu.cn}

\affil[1]{\orgdiv{Information Hub}, \orgname{The Hong Kong University of Science and Technology (Guangzhou)}, \orgaddress{ \city{Guangzhou}, \country{China}}}

\affil[2]{\orgdiv{Function Hub}, \orgname{The Hong Kong University of Science and Technology (Guangzhou)}, \orgaddress{ \city{Guangzhou}, \country{China}}}

\affil[3]{\orgdiv{Guangzhou Municipal Key Laboratory of Materials Informatics}, \orgaddress{ \city{Guangzhou}, \country{China}}}

\affil[4]{\orgdiv{Materials Genome Institute, Shanghai University}, \orgaddress{ \city{Shanghai}, \country{China}}}

\abstract{
Perovskite solar cells (PSCs) have been considered as a next-generation disruptive photovoltaic technology, yet their advancement is constrained by the complexity of perovskite recipe with high-dimensional material and process design space. 
Despite the impressive general reasoning of Large Language Models (LLMs), they struggle with two limitations for application in PSCs: an inability to align general semantics with the perovskite domain knowledge, and an inefficiency in navigating high-dimensional perovskite material and recipe design spaces. 
To address these limitations, we introduce a domain-knowledge-guided framework PVK-LLM, a specialized model to serve as an expert to bridge general semantics with perovskite domain knowledge. 
By integrating this domain knowledge into a hierarchical Bayesian Optimization workflow, our approach efficiently navigates the high-dimension design space on a solar cell simulator platform.
The domain knowledge resolves cold-start problems while dynamically adapting to simulator feedback.
Moreover, in an individual wet-lab experiment aimed at maximizing power conversion efficiency (PCE), our framework autonomously proposes a novel synergistic four-component recipe comprising specialized organic passivation recipe (3MTPAI, PDAI$_2$, EDAI$_2$, and PipDI) which has not been reported in existing literature.
This AI-designed recipe effectively achieves a champion PCE value of over $26.0 \%$, approaching world records achieved through extensive expert trial-and-error. 
Our approach can effectively enable LLM comprehend the domain knowledge, which can efficiently navigate in a high-dimensional, capable to accelerate the advancement in real-world perovskite as well as other material science development.

}




\maketitle

\section{Introduction}

Perovskite solar cells (PSCs) stand at the forefront of next-generation photovoltaics~\cite{green2014emergence,yang2025towards,zou2025electrically,kojima2009organometal}, with efficiencies now exceeding $27\%$~\cite{xiong2025homogenized}. 
However, unlocking their full potential requires navigating a vast, high-dimensional design space that spans precursor compositions, solvent engineering, and processing parameters~\cite{song2022spring,cakan2025bayesian,chen2020bayesian,liu2022machine}. 
Current paradigms relying on researcher-driven trial-and-error are inherently laborious and inefficient, struggling to identify global optima within this complex combinatorial landscape~\cite{goetz2022challenge,miret2025enabling}.

While Large Language Models (LLMs) offer a pathway to leverage textual knowledge~\cite{yin2024survey,weiemergent,mirza2025framework,he2025generalized}, existing LLMs ChemCrow~\cite{m2024augmenting}, Coscientist~\cite{boiko2023autonomous}, SciToolAgent~\cite{ding2025scitoolagent}, Chemma~\cite{zhang2025large}, and GPT Series~\cite{jablonka2024leveraging} encounter limitations when applied to the specialized domain of PSCs.
A critical gap remains between the qualitative semantic capabilities of existing LLMs and the precise, quantitative demands of perovskite engineering~\cite{alampara2025probing}. 
Existing LLMs often offer scientifically invalid recipes due to a lack of domain knowledge and fail to perform quantitative extrapolation in multi-parametric design spaces. 
Consequently, they struggle to orchestrate the complex, closed-loop optimization required for device breakthroughs.

To address these limitations, we present a systematic active learning framework that synergizes domain knowledge with autonomous recipe design. 
We propose PVK-LLM, a domain-knowledge-guided LLM engineered to bridge the semantic gap between unstructured scientific literature and precise materials recipes and experimental parameters for PSC research.
PVK-LLM is established through a three-stage curriculum learning strategy designed to simulate the training trajectory of a domain expert from knowledge alignment to quantitative interpretation. 
Furthermore, to navigate the high-dimensional design space, we construct a hierarchical Bayesian Optimization workflow driven by PVK-LLM (PVK-BO). 
Unlike traditional methods that search blindly, PVK-BO leverages internalized domain knowledge to narrow the search space effectively, addressing the cold-start problems and iteratively refining recipes based on real-time feedback from experiment verification.

Our model achieves state-of-the-art accuracy on domain-specific benchmarks and outperforms existing methods in simulator recipe design tasks.
In the closed-loop wet-lab experiment, we demonstrate the real-world applicability of our framework through designing the high-power conversion efficiency (PCE) p-i-n PSCs autonomously.
Without human guidance, PVK-LLM identifies interface passivation as the bottleneck and autonomously designs a novel, previously unreported recipe with precise mixture ratios of a four-component synergistic passivation system (3MTPAI, PDAI$_2$, EDAI$_2$, and PipDI). 
The novel recipe achieves a champion PCE of $26.0 \%$, approaching world records typically achieved through extensive expert experience~\cite{xiong2025homogenized}. 
These results reveal that domain-knowledge-guided AI can not only internalize scientific expertise but also accelerate advancement in real-world science challenges.
The versatility of our domain-knowledge-driven framework enables extension to other chemistry or material domains, including battery electrolyte optimization and organic photovoltaic light-active layer screening.

\section{Results}

\subsection{Overview of Framework}
\begin{figure*}
    \centering
    \includegraphics[width=1.0\linewidth]{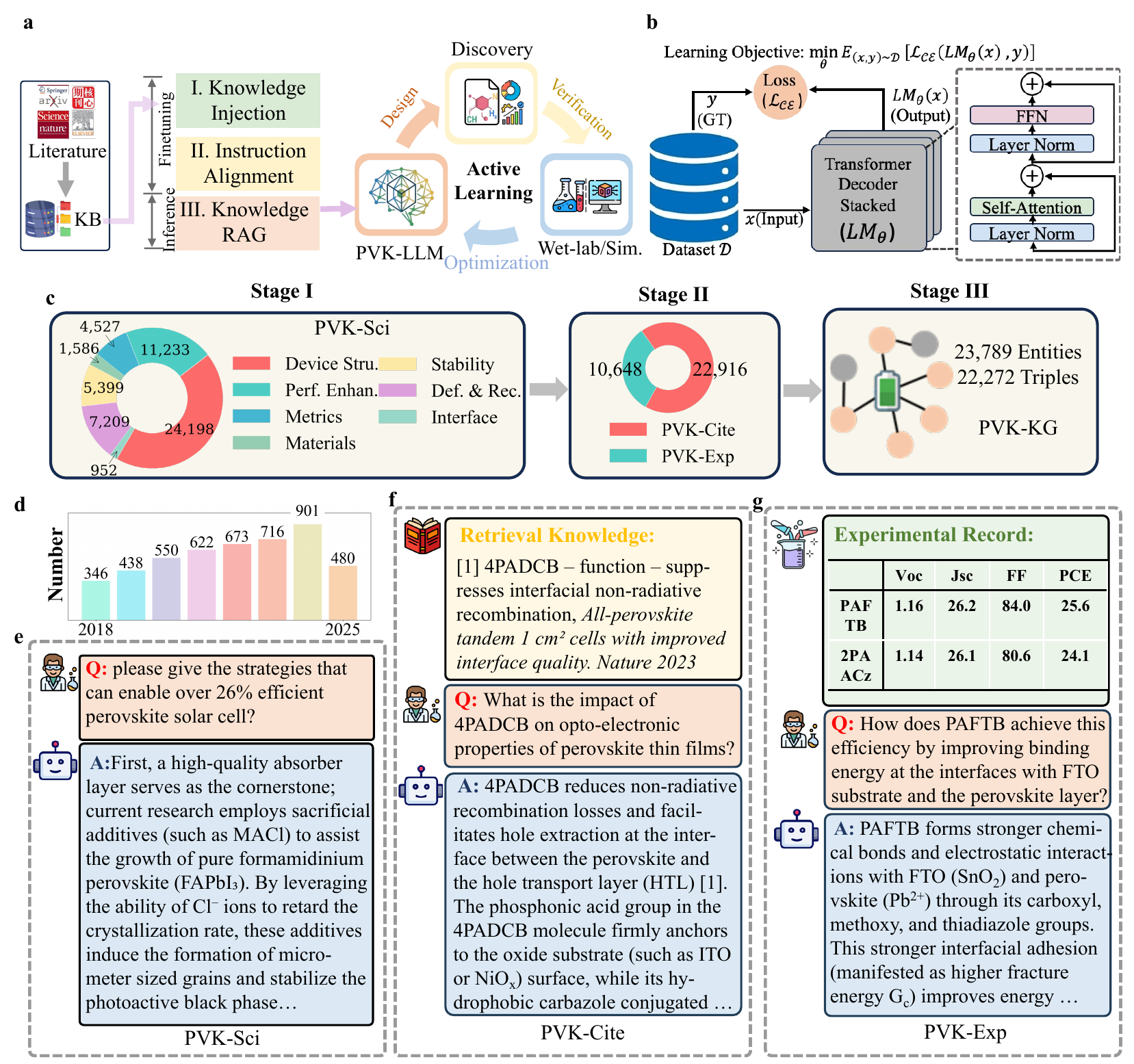}
    \caption{Architecture and implementation of the PVK-LLM framework for autonomous scientific advancement. 
    (a) Overview of the closed-loop active learning workflow to integrate domain knowledge. 
    Literal knowledge extracted from scientific literature constructs a knowledge base (KB) to fine-tune the model through three phases. 
    This enables PVK-LLM to drive the design, verification, and optimization cycles in wet-lab and simulator (sim.) environments. 
    (b) The model architecture and learning objective for PVK-LLM.
    GT means ground truth.
    (c) Composition of the three-stage curriculum learning datasets. 
    Stage I (PVK-Sci) covers domain knowledge; Stage II integrates knowledge grounding (PVK-Cite) and experiment analysis data (PVK-Exp); Stage III establishes a structured knowledge graph (PVK-KG).
    Performance Enhancement and Defect \& Recombination are abbreviated as Perf. Enhan. and Def. \& Rec., respectively.
    (d) The distribution of the perovskite literature corpus, spanning from $2018$ to $2025$ July. 
    (e-g) Representative examples of the instruction-tuning datasets: 
    (e) PVK-Sci addresses scientific QA; 
    (f) PVK-Cite incorporates retrieval-based evidence for knowledge grounding; 
    (g) PVK-Exp utilizes structured experimental records to facilitate detailed mechanism analysis and performance attribution.}
    \label{fig:overview}
\end{figure*}

In this work, we present a systematic framework designed to accelerate scientific advancement in the field of PSCs through closed-loop active learning, as illustrated in Fig.~\ref{fig:overview}. 
At the core of this framework lies PVK-LLM, which serves as a domain knowledge engine to bridge the semantic gap between unstructured scientific literature and precise materials recipes and experimental parameters for PSC research.
It is established by injecting domain knowledge via a three-stage curriculum learning strategy (Fig.~\ref{fig:overview} (a)).
The fundamental architecture of PVK-LLM utilizes a stacked Transformer decoder optimized via cross-entropy loss, illustrated in Fig.~\ref{fig:overview} (b).

The foundation of PVK-LLM is a comprehensive corpus comprising over $4,000$ high-impact articles published from $2018$ to July $2025$ (Fig.~\ref{fig:overview} (d)), from which we curate three specialized datasets (Fig.~\ref{fig:overview} (c)).
In Stage I (Knowledge Injection), we utilize the Qwen2.5-32B model as our foundational backbone and perform fine-tuning on PVK-Sci. 
This instruction dataset PVK-Sci comprises a total of $55,104$ question-answering (QA) pairs, curated to cover seven research themes detailed in Fig.~\ref{fig:supp_1} (b): Device Structure, Performance Enhancement, Interface Engineering, Stability, Materials, Defects \& Recombination, and Metrics. 
By internalizing these diverse knowledge forms, the model mitigates the gap between the general knowledge and domain knowledge for perovskite research. 
The Fig.~\ref{fig:overview} (e) displays representative data samples from this stage, showing how specific additives (e.g., MACI) are correlated with crystallization kinetics.
In Stage II (Instruction Alignment), we conduct continual fine-tuning on the Stage I model based on two datasets: PVK-Cite and PVK-Exp.
We introduce the Perovskite Knowledge Citation dataset (PVK-Cite), consisting of $22,916$ QA pairs, which trains the model to ground its knowledge in specific literature citations. 
For instance, as shown in the Fig.~\ref{fig:overview} (f), the data sample demonstrates how the model supports its answer regarding the 4PADCB molecule by explicitly citing its specific function in suppressing non-radiative recombination. 
Perovskite Experimental Analysis dataset (PVK-Exp), comprising $10,648$ QA pairs, equips the model with the ability to interpret quantitative experimental records. 
As shown in the sample in the right panel of Fig.~\ref{fig:overview} (g), this allows the model to accurately parse performance metrics from tables and analyze efficiency mechanisms.
Finally, Stage III implements a Retrieval-Augmented Generation (RAG) mechanism to bridge the static parameters of the LLM with dynamic scientific updates. 
This is achieved through a self-updating Perovskite Knowledge Graph (PVK-KG) managed by an automated pipeline (Fig.~\ref{fig:kg_supp2} (a)) guided by an expert-defined schema (Fig.~\ref{fig:supp_1} (b)).
Our PVK-KG contains $23,789$ entities and $22,272$ triples.
This pipeline continuously processes up-to-date literature to organize structured entities and triples, ensuring the model mediates between its internal representations and the latest scientific discoveries without the need for frequent retraining.

Building upon this domain knowledge, the framework operates through a dynamic, closed-loop Active Learning cycle (Fig.~\ref{fig:overview} (a)). 
In this loop, the model navigates the high-dimensional material-recipe and experimental-parameter space for PSCs. 
It utilizes its domain knowledge to design initial experiments, proposes specific recipes for Verification in simulator or wet-lab experiments separately, and subsequently refines its strategies through Optimization based on feedback.
Specifically, this iterative refinement is powered by our proposed PVK-BO algorithm (detailed in Sec.~\ref{sec:pvk_bo_algorithm}). 
PVK-BO leverages the internalized domain knowledge of PVK-LLM for optimization of material properties for targeted functional layers, such as charge-transport and passivation layers. 
To validate its effectiveness, we first conducted functional-layer property optimizations using feedback from physics-based device simulators. 
In another independent wet-lab experiment, it also performs recipe design by recommending passivation molecules with precisely designed composition ratios.
By integrating the extrapolation capability of LLMs with the exploitation efficiency of Bayesian optimization methods, this approach enables the agent to efficiently locate optimal material compositions within a search space, accelerating the experimental advancement process compared to traditional trial-and-error methods.
We evaluate our framework from three different independent approaches: the benchmark experiment (Sec.~\ref{sec:performance_of_benchmark_data}), the simulator experiment (Sec.~\ref{sec:performance_of_simulation_experiment}), and the wet-lab experiment (Sec.~\ref{sec:performance_of_wet-lab_experiment}).

\subsection{Performance of Benchmark}
\label{sec:performance_of_benchmark_data}
To evaluate the domain capabilities of PVK-LLM, we conduct comprehensive assessments focusing on two datasets: PVK-MCQ and PVK-QA.

% For the benchmark evaluation, we design the evaluation dataset, including multiple-choice questions and question answering in the perovskite domain. 
% The evaluation dataset is also extracted from the top-level publications in the perovskite domain with our multi-agent framework and an extra expert double check. The evaluation dataset contains $1,103$ question answering named \textbf{PVK-QA} and $1,103$ multiple choice questions named \textbf{PVK-MCQ}.
% For question answering, we set the LLM-as-a-Judge~\cite{zheng2023judging} score and a human expert judge as the evaluation metric. 
% In our experiments, we find that both metrics can effectively measure the quality of question answering and consistency with each other.
% For multiple-choice questions, we set the accuracy as the evaluation metric. 
\subsubsection{Benchmark Design}
To quantitatively evaluate the performance of PVK-LLM in the specialized perovskite domain, we curate an evaluation benchmark derived from publications in the field's top-tier journals. 
The benchmark dataset comprises a total of $2,206$ samples, divided into two specific parts: PVK-QA, which includes $1,103$ open-ended question-answering pairs, and PVK-MCQ, consisting of $1,103$ multiple-choice questions. 
Regarding evaluation metrics, we employ Accuracy for the PVK-MCQ tasks. 
For the open-ended generation tasks in PVK-QA, we adopt a dual-evaluation mechanism utilizing both LLM-as-a-Judge~\cite{zheng2023judging} and human expert scoring.
To ensure a granular and holistic assessment, we evaluate the responses across five specific dimensions: Accuracy, Completeness, Relevance, Clarity, and Overall Quality.
The prompt in the evaluation pipeline is shown in Fig.~\ref{fig:prompt} (b).

\subsubsection{Benchmark Evaluation}
\label{sec:benchmark_evaluation}
\begin{figure*}
    \centering
    \includegraphics[width=1.0\linewidth]{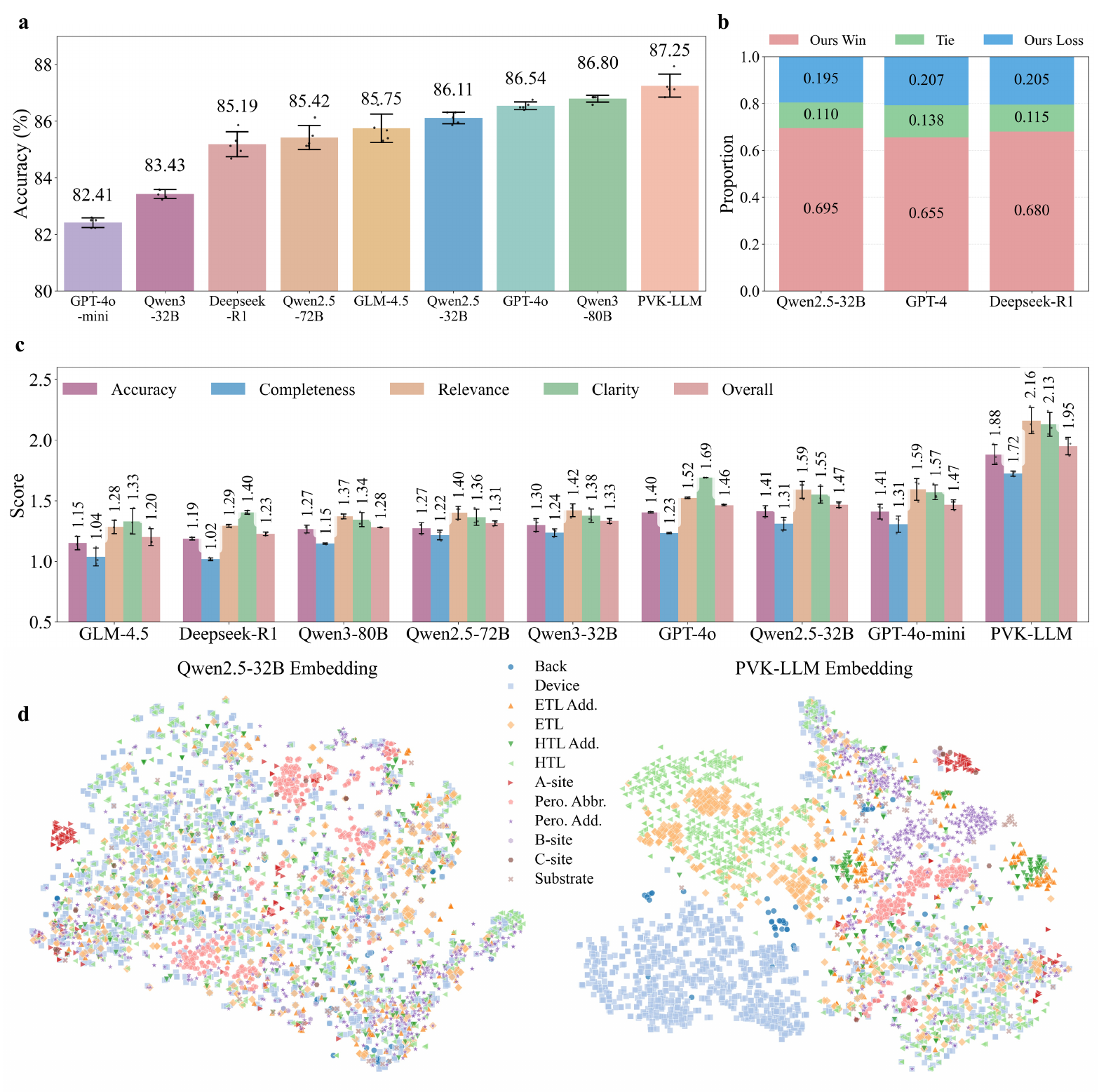}
    \caption{Benchmark results in domain-specific evaluation. 
    (a) Objective evaluation on the PVK-MCQ dataset. PVK-LLM achieves a state-of-the-art score of $87.25 \%$, surpassing powerful general LLMs, demonstrating the efficacy of domain knowledge injection. 
    All experiments are conducted with five independent runs.
    (b) Subjective Human Expert evaluation. Expert reviewers rate PVK-LLM higher than other models. 
    (c) Detailed metric breakdown using LLM-as-a-Judge. 
    Our PVK-LLM outperforms baselines across accuracy, completeness, relevance, and clarity. 
    All experiments are conducted with five independent runs.
    (d) t-SNE visualization of embedding spaces. Comparison between the base model (Qwen2.5-32B) and PVK-LLM reveals that the fine-tuned model forms more distinct, semantically meaningful clusters of perovskite materials. 
    The legend abbreviations correspond to specific material categories: Back (Back Contact Layer), Device (Complete Device Structure), ETL/HTL (Electron/Hole Transport Layers) and their additives (Add.), A/B/C-site (Perovskite Cations), Pero. Add./Abbr. (Perovskite Additives/Abbreviations), and Substrate.
    }
    \label{fig:benchamark_exp}
\end{figure*}

We evaluate PVK-LLM as illustrated in Fig.~\ref{fig:benchamark_exp}. 
To ensure robustness, all experiments are performed in a zero-shot setting, with results averaged over five independent runs.
First, we assess objective domain knowledge using the PVK-MCQ benchmark (Fig.~\ref{fig:benchamark_exp} (a)). 
PVK-LLM achieves a state-of-the-art accuracy of $87.25$, surpassing general LLMs such as Qwen2.5-72B and GPT-4o. 
This performance demonstrates that our domain-specific fine-tuning through our curricular learning injects perovskite domain knowledge into the model.
Second, for the open-ended generation tasks on PVK-QA, we employ an LLM-Judge to evaluate performance across five metrics (Fig.~\ref{fig:benchamark_exp} (c)): accuracy, completeness, relevance, clarity, and overall score. 
PVK-LLM outperforms all baseline models. 
The performance in relevance and completeness indicates its ability to generate comprehensive scientific explanations that align closely with user intent.
Finally, to validate practical utility, we employ five experts to conduct a pairwise Human Evaluation against representative strong baselines (Fig.~\ref{fig:benchamark_exp} (b)).
The results indicate that PVK-LLM achieves a dominant win rate across all comparisons: $69.5\%$ against Qwen2.5-32B, $65.5\%$ against GPT-4, and $68.0\%$ against Deepseek-R1.
This preference margin highlights that while general LLMs may produce fluent text, they often lack the depth and reliability required by domain experts, whereas PVK-LLM meets the precise standards of scientific research.

Furthermore, to evaluate the underlying structure of this internalized domain knowledge in PVK-LLM, we visualize the semantic representations of perovskite entities using t-SNE (Fig.~\ref{fig:benchamark_exp} (d)). 
We extract the over $7,500$ perovskite materials from the existing database~\cite{jacobsson2022open}.
We visualize the high-dimensional embeddings of various perovskite-related concepts from both the baseline Qwen2.5-32B and our PVK-LLM. 
As shown in the left panel, the embeddings from the Qwen2.5-32B exhibit a random distribution where functional categories are entangled. 
This suggests that the base model lacks a structured understanding of domain knowledge taxonomy. 
In contrast, the embedding space of PVK-LLM (right panel) demonstrates clustering structures. 
Concepts belonging to the same category (e.g., Hole Transport Layers shown in green) are grouped, while boundaries between different functional layers are well-separated. 
This semantic separation confirms that our fine-tuning process through our curricular learning has aligned the model's latent space with expert domain knowledge, enabling it to discriminate between complex scientific concepts rather than mimicking text patterns.

\subsection{Adaptive Performance with a Simulator}
\label{sec:performance_of_simulation_experiment}
To validate the efficient navigation capability of PVK-BO within high-dimensional parameter spaces, we conduct experiments within the simulator focusing on two independent tasks in perovskite device design: band alignment and doping optimization of carrier transportation layers.

\subsubsection{Experimental Setup and Framework}
\begin{figure*}
    \centering
    \includegraphics[width=1.0\linewidth]{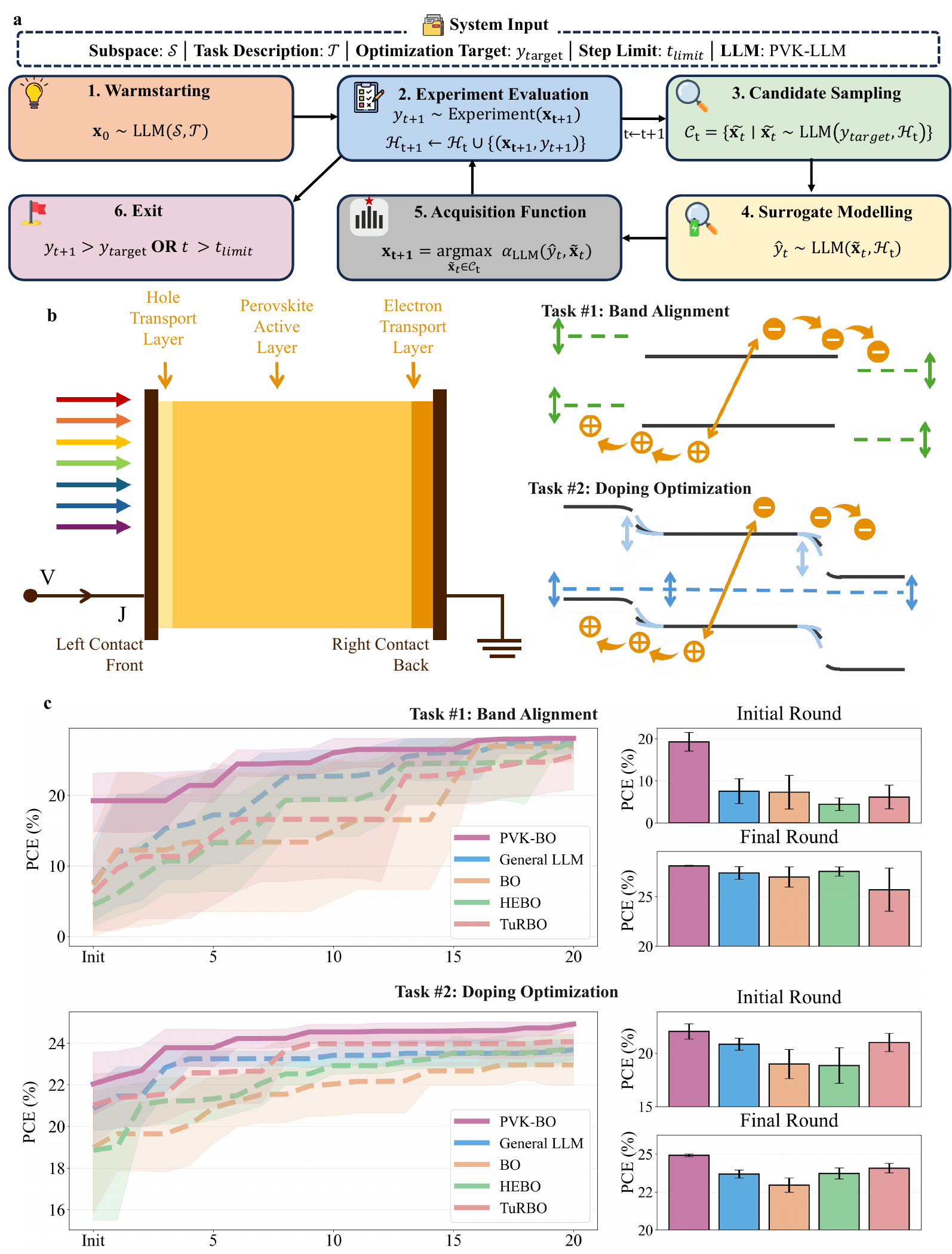}
    \caption{Framework and performance of simulator experiments. 
    (a) Workflow of the PVK-LLM-based Bayesian Optimization (PVK-BO).
    (b) Schematics of the physical optimization tasks. 
    Task \#1 optimizes Band Alignment to minimize energy barriers at interfaces, while Task \#2 optimizes Doping concentrations to align Fermi levels and reduce defects. 
    (c) For Band Alignment and Doping Optimization, PVK-BO achieves maximum PCEs of $26.52 \%$ and $25.44 \%$, respectively. All experiments are conducted with five independent runs.}
    
    \label{fig:sim_exp}
\end{figure*}

We employ the PVK-LLM-based Bayesian Optimization (PVK-BO), an optimization framework that leverages the generative capabilities of PVK-LLM. 
For semiconductor physics validation, we incorporate the Solar Cell Capacitance Simulator (SCAPS-1D) into the optimization loop.
As illustrated in Fig.~\ref{fig:sim_exp} (a), this workflow operates through five key stages:
(1) Warmstarting, where the LLM provides high-quality initial candidates; 
(2) Recipe Design, generating new candidate samples; 
(3) Surrogate Modelling, where the LLM acts as a probabilistic model to predict performance; 
(4) Acquisition Function, selecting the most promising recipe to maximize expected gains;
(5) Experiment Evaluation, where the recipe is validated via simulator to update the dataset.

The simulator experiments are conducted by using a perovskite device structure within SCAPS-1D~\cite{burgelman2000modelling} (Fig.~\ref{fig:sim_exp} (b)). 
The primary evaluation metric is PCE, which quantifies the device's ability to convert incident light into electrical energy. 
We focus on two optimization tasks:
\begin{enumerate}
\item \textbf{Band Alignment.}
As shown in Fig.~\ref{fig:sim_exp} (b) Task \#1, this task tunes the electron affinity and band gap of the perovskite, Electron Transport Layer (ETL), and Hole Transport Layer (HTL). 
The goal is to minimize energy barriers at the Conduction Band Minimum (CBM) and Valence Band Maximum (VBM) to facilitate efficient carrier extraction and suppress interfacial recombination.
\item \textbf{Doping Optimization.}
Depicted in Task \#2, this task adjusts the acceptor and donor doping concentrations in the transport layers. 
Doping determines the Fermi level of the transport layer, the band bending of PVK near the transport layer, and the conductivity of the transport layer. Proper doping could facilitate the carrier extraction and reduce the interface recombination velocity.
\end{enumerate}

\subsubsection{Experimental Results}
We evaluate the performance of PVK-BO against several baselines, utilizing both optimization trajectories and distributional histograms as shown in Fig.~\ref{fig:sim_exp} (c). 
Our experiments are repeated $5$ times.
The baselines include optimization based on Qwen2.5-32B (General LLM) and three machine learning optimization methods: Standard BO~\cite{liu2025disentangling}, HEBO~\cite{cowen2022hebo}, and TuRBO~\cite{eriksson2019scalable}.
In Fig.~\ref{fig:sim_exp} (c), the solid curves track the mean PCE aggregated across five independent runs, with the shaded regions representing the variance distribution, which indicates the stability of each algorithm. 
The bar charts on the right detail the PCE distribution at critical stages: the Initial Round illustrates the distribution of initialized values to assess warm-start capabilities, while the Final Round depicts the performance at the $20$-th iteration to evaluate the convergence.

For Task \#1 (Band Alignment), the results indicate the influence of domain knowledge on initialization.
The Initial Round histogram shows a difference in starting points. 
PVK-BO achieves a higher initial mean PCE than Qwen2.5-32B and other algorithmic baselines. 
This distribution suggests that while general LLMs possess broad reasoning capabilities, the specialized training of PVK-LLM provides more precise initial parameters based on the domain knowledge.
Even compared to baselines like HEBO and TuRBO, our method demonstrates an advantage in initialization efficiency, effectively bypassing the initial exploration phase that constrains these data-driven algorithms.
Although the trajectory shows that advanced algorithms like HEBO improve after multiple iterations, PVK-BO maintains a higher efficiency with minimal variance throughout the process.

In Task \#2 (Doping Optimization), the comparison further differentiates general and domain knowledge.
The Final Round histogram shows differences in the optimization ceiling. 
While Qwen2.5-32B plateaus, PVK-BO reaches a final PCE of $25.44\%$.
While HEBO and TuRBO outperform standard BO, they still yield higher variance and slightly lower final efficiency than PVK-BO, underscoring the value of domain knowledge in stabilizing the search for global optima.
Furthermore, the final histogram shows that PVK-BO has the lowest variance, indicating stable performance. 
These results from two independent tasks suggest that integrating domain knowledge assists the framework in locating the global optimum, outperforming methods that rely on general knowledge or standard algorithms.

\subsection{Adaptive Performance with Wet-lab Experiment}
\label{sec:performance_of_wet-lab_experiment}

To validate the practical applicability of our framework beyond simulator environments, we deploy PVK-LLM in another independent physical wet-lab setting to guide the closed-loop fabrication and optimization of high-efficiency perovskite solar cells, as shown in Fig.~\ref{fig:wet-lab}.
The details of the recipe can be seen in Sec.~\ref{sec:device_fabrication_in_wet_lab}.

\subsubsection{Experimental Design}

We establish an iterative optimization experiment targeting normal-bandgap ($\sim1.5eV$) devices with high PCEs (Fig.~\ref{fig:wet-lab} (a)). 
The workflow is initialized by utilizing PVK-LLM to define a complete recipe for the specified device architecture (Glass/FTO/HTL/Perovskite/Passivation/SnO$_2$/Cu). 
Then this initial wet-lab executes the defined recipe and obtains the epoch $0$ experiment results, including both the photovoltaic performance and fabrication parameters of perovskite solar cells.
In the subsequent screening step, assimilating these experimental data allows PVK-LLM to diagnose performance bottlenecks within the baseline recipe and formulate optimization strategies. 
Consequently, the optimization phase integrates the PVK-BO framework detailed in Sec.~\ref{sec:optimization_application} to navigate the high-dimensional parameter space.

% To validate the practical efficacy and closed-loop optimization capability of PVK-LLM in guiding real-world perovskite solar cell (PSC) fabrication, we conduct an iterative optimization experiment focusing on fabricating normal-bandgap perovskite solar cells with the explicit goal of achieving high PCE approaching $26\%$. 
% This process assess whether the model could autonomously diagnose performance bottlenecks from experimental data and iteratively optimize device fabrication strategies toward this target. 
% We initiate the process by tasking PVK-LLM to generate a complete recipe for a single-junction normal-bandgap PSC (device structure: Glass/FTO/HTL/Perovskite/Passivation/SnO$_2$/Cu).
% Upon obtaining the Epoch 0 dataset—including photovoltaic parameters and fabrication details—we input the data to PVK-LLM for autonomous analysis. 
% The model identified the key performance-limiting component and proposed optimized strategies. 
% To navigate the high-dimensional optimized parameter space, we deploy PVK-LLM within a Bayesian optimization (PVK-BO) framework.

\subsubsection{Experimental Results}
\begin{figure*}
    \centering
    \includegraphics[width=1.0\linewidth]{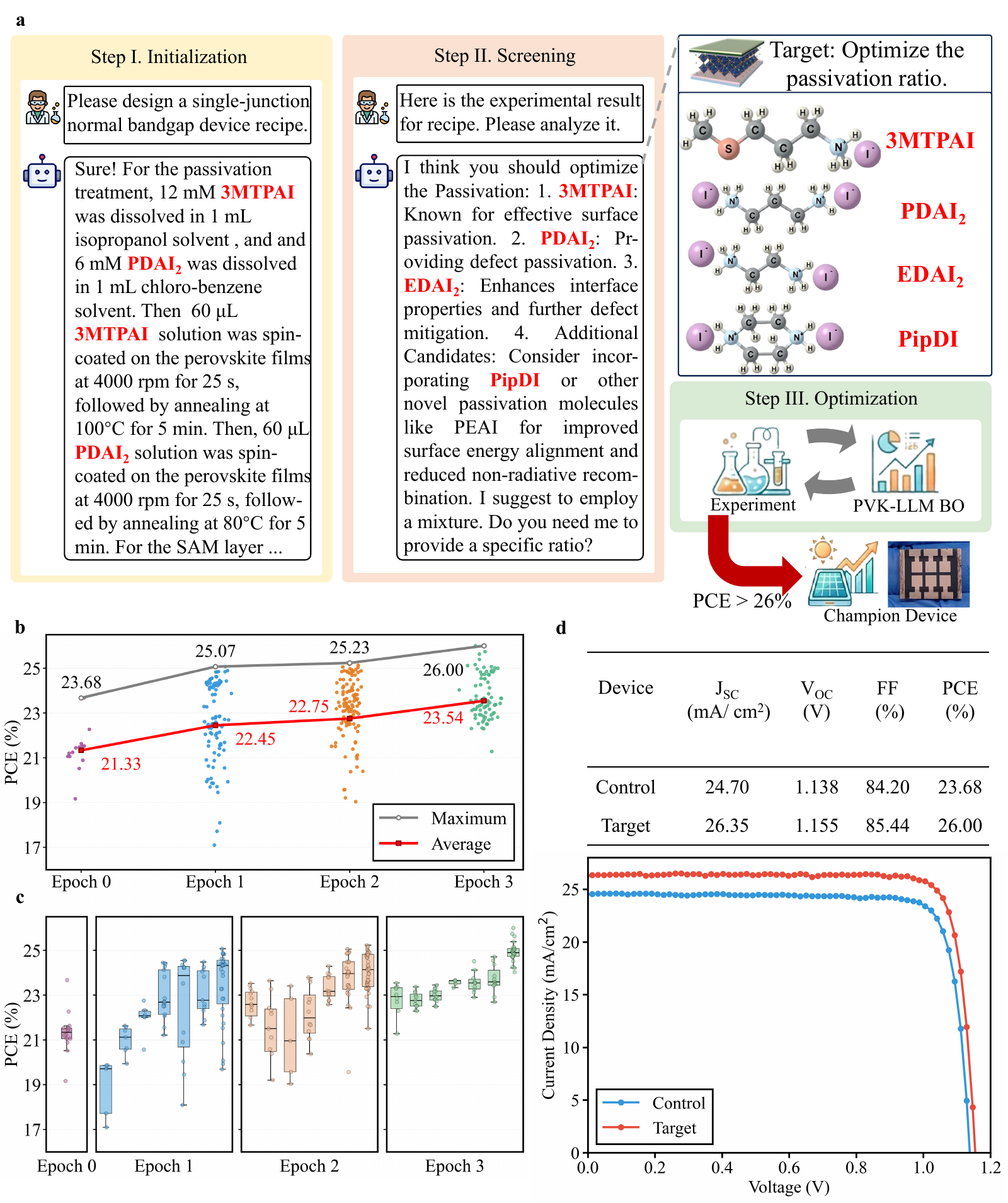}
    \caption{Wet-lab experimental validation and optimization of perovskite solar cells empowered by PVK-LLM.
    (a) The human-in-the-loop active learning workflow. The process proceeds through three stages: Initialization of the base recipe, Screening of recipe based on experimental feedback, and Optimization of the mixture ratio using PVK-BO.
    (b) Evolution of the PCE distribution across iterative experimental epochs.
    The results demonstrate a clear upward trajectory in device performance from Epoch 0 to Epoch 3.
    (c) PCE distribution for each recipe from Epoch 0 to Epoch 3. 
    We let our frame work to recommend seven recipes in each epoch.
    (d) Current density-voltage (J-V) characteristics of the champion device (``Target") compared to the control group.
    The optimized device achieves a performance boost with a peak PCE of $26.00 \%$.
    }
    \label{fig:wet-lab}
\end{figure*}

To validate the practical effectiveness of our framework, we conduct an independent closed-loop wet-lab optimization to enhance the performance of p-i-n perovskite solar cells. 
In addition to the PCE value, we assess the recipe through detailed performance parameters such as: Open-Circuit Voltage (V$_{OC}$), Short-Circuit Current Density (J$_{SC}$), and Fill Factor (FF).
As illustrated in the human-in-the-loop workflow (Fig.~\ref{fig:wet-lab} (a)), the process begins with an initialization stage (Epoch $0$), where a standard device recipe served as the baseline. 
The champion device from this epoch achieved a PCE of $23.68\%$ (V$_{OC} = 1.138$ V, J$_{SC} = 24.70$ mA/cm$^2$, FF = $84.20\%$).
In the screening stage, the comprehensive dataset from Epoch $0$ is analyzed by PVK-LLM. 
Leveraging its domain knowledge, the model identifies the passivation layer as the primary bottleneck for further efficiency gains. 
Moving beyond initial dual-layer passivation methods, PVK-LLM proposes a synergistic four-component system integrating 3MTPAI (surface passivation), PDAI$_2$ (defect passivation), EDAI$_2$ (interface enhancement), and PipDI (energy level alignment).
The optimization of this high-dimensional composition space was then navigated using the PVK-BO framework (Fig.~\ref{fig:wet-lab} (a), Step III). 
The iterative refinement proceeds as follows:
\begin{enumerate}
    \item In Epoch 1, based on the model's initial recommended mixture (3MTPAI : PDAI$_2$ : EDAI$_2$ : PipDI = $0.20:0.55:0.10:0.15$), the champion PCE rise to $25.07\%$.
    \item In Epoch 2, after feeding these results back into the loop, a refined mixture (3MTPAI : PDAI$_2$ : EDAI$_2$ : PipDI = $0.35:0.20:0.35:0.15$) further improves the champion PCE to $25.23\%$.
    \item In Epoch 3, the model recommends the passivation ratio (3MTPAI : PDAI$_2$ : EDAI$_2$ : PipDI = $0.45:0.15:0.35:0.05$), yielding a champion PCE of $26.00\%$ (Voc = $1.155$ V, Jsc = $26.35$ mA/cm$^2$, FF = $85.44 \%$). 
\end{enumerate}

The iterative evolution of the experiments is illustrated in Fig.~\ref{fig:wet-lab} (b) and (c), where the upward trajectory of both maximum and average PCE confirms the effectiveness of the closed-loop optimization.
Epoch 0 corresponds to the Initialization Phase (Step I), where PVK-LLM proposes a single starting recipe for a two-component passivation system. 
Subsequently, Epochs 1 through 3 represent the Optimization Phase (Step III). 
In each of these epochs, PVK-LLM recommends seven distinct composition ratios for the four-component passivation system.
The PCE for each recipe from Epoch 0 to Epoch 3 exhibits a narrowing variance, showing the model's increasing precision in locating the optimal space.
This wet-lab experiment demonstrates that PVK-LLM can function as a powerful autonomous engine, driving complex experimental optimizations to reach high-efficiency targets ($26\%$) in iterations.
As shown in the J-V characteristics (Fig.~\ref{fig:wet-lab} (d)), the target device optimized by PVK-LLM exhibits a holistic performance boost compared to the ``Control" group.

\subsection{Analysis on PVK-LLM and PVK-KG}
In this section, we analyze and validate more studies of PVK-LLM and PVK-KG through experiments, as shown in Fig~\ref{fig:ablation_comparison_AB}.

\subsubsection{Analysis on Curricular Learning}
\begin{figure*}
    \centering
    \includegraphics[width=1.0\linewidth]{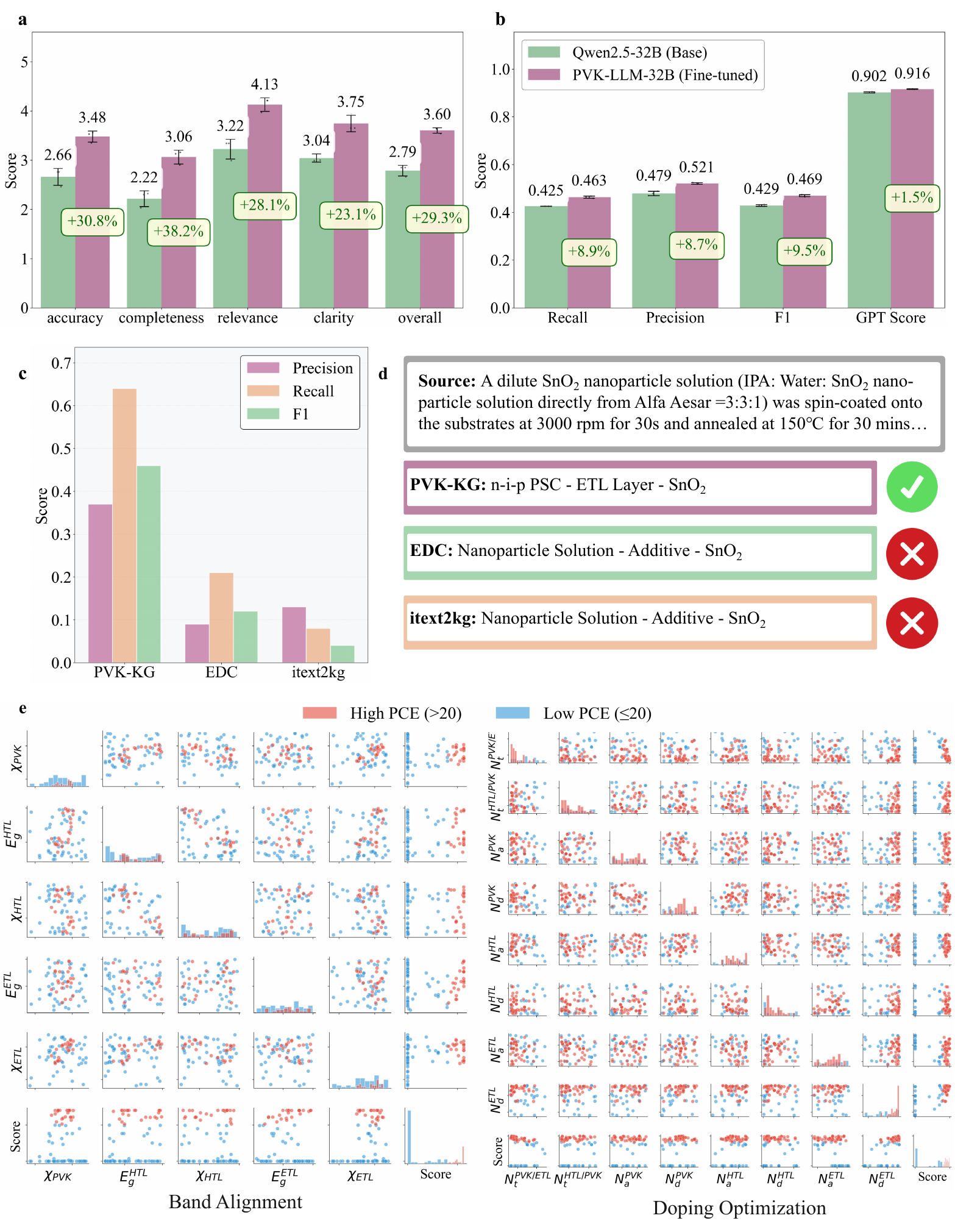}
    \caption{Other analysis of our pipeline. 
    (a) and (b) are conducted with five independent runs.
    (a) Model ablation on experiment analysis. 
    Comparing the PVK-LLM (Fine-tuned) with Qwen2.5-32B (Base) shows gains. 
    (b) Evaluation of knowledge grounding. The PVK-LLM (Fine-tuned) demonstrates improvement.
    (c) Assessment of PVK-KG quality. 
    PVK-KG achieves higher metrics compared to other automatic methods. 
    (d) Case study on KG construction. 
    Our pipeline identifies SnO$_2$ as the ETL layer while other methods fail. 
    (e) The correlation of simulator experiment parameters.
    }
    \label{fig:ablation_comparison_AB}
\end{figure*}

\noindent$\bullet$ \textbf{Experiment Setting}.
To validate the effectiveness of our Stage II fine-tuning strategy, we conduct a comparative analysis between PVK-LLM (Fine-tuned) and its foundational base model, Qwen2.5-32B (Base), as shown in Fig.~\ref{fig:ablation_comparison_AB} (a) and (b).
Since Qwen2.5-32B serves as our backbone, we use it as a baseline to demonstrate the effectiveness of our Stage II training.
The evaluation utilizes unseen samples from the held-out testsets of our PVK-Exp and PVK-Cite datasets.
To ensure the assessment, we employ different metric systems for each task.
For the experiment analysis task (PVK-Exp test set), we adopt the comprehensive five-dimensional metric system established in our main benchmark (Sec.~\ref{sec:benchmark_evaluation}).
For the knowledge grounding task (PVK-Cite test set), we utilize standard retrieval metrics, where Recall measures the breadth of knowledge coverage, Precision reflects the trustworthiness of the generated citations, and the F$_1$ score serves as a balanced measure of overall performance.

\noindent$\bullet$ \textbf{Results}.
The comparative results highlight the impact of curricular learning.
In the task of PVK-Exp test set (Fig.~\ref{fig:ablation_comparison_AB} (a)), PVK-LLM (Fine-tuned) demonstrates a performance gain, achieving a $38.2\%$ increase in completeness and a $30.8\%$ improvement in accuracy.
While the Qwen2.5-32B (Base) tends to provide general descriptions, PVK-LLM exhibits the specific domain knowledge necessary to identify experimental features and deliver scientific analyses.
This advantage extends to the PVK-Cite test set(Fig.~\ref{fig:ablation_comparison_AB} (b)).
PVK-LLM shows the gains in F$_1$ score ($+9.5\%$) and Precision ($+8.7\%$).
These results confirm that our fine-tuning aligns the model's internal representations with established \textbf{domain knowledge}, thereby ensuring that generated insights are grounded in academic facts.

\subsubsection{Analysis on PVK-KG Construction}

\noindent$\bullet$ \textbf{Experiment Setting}.
We construct PVK-KG based on our proposed pipeline (in Fig~\ref{fig:kg_supp2} (a)).
We evaluate our pipeline through a comparative analysis against existing automated KG construction methods, specifically EDC~\cite{zhang2024extract} and itext2kg~\cite{lairgi2024itext2kg} (Fig.~\ref{fig:ablation_comparison_AB} (c)).
For this assessment, we curate a dataset of $50$ representative papers, where knowledge triples were manually annotated by domain experts to serve as the ground truth.
We quantify the performance by comparing the model-extracted knowledge against this ground truth using three standard metrics:
Precision assesses the correctness of the extracted triples, Recall measures the comprehensiveness of knowledge coverage, and the F$_1$ score provides a balanced metric of overall extraction quality.

\noindent$\bullet$ \textbf{Results}.
The results demonstrate an advantage for our approach.
While baseline models exhibit limited recall, our pipeline achieves a superior extraction of knowledge entries by leveraging the context of literature.
The performance gap primarily stems from the baselines' inability to synthesize the global literature context.
Existing methods often rely on local semantic proximity, leading to erroneous classifications when a material's role is implied rather than explicitly stated.
This limitation is illustrated in the SnO$_2$ extraction case study (Fig.~\ref{fig:ablation_comparison_AB} (d)).
Lacking an architectural understanding of PSCs, baselines incorrectly categorize SnO$_2$ as a generic ``additive" simply due to its description within a solution-processing context.
In contrast, our baseline correctly identifies its functional role as the ETL.
This accuracy is achieved by our pipeline's ability to interpret the whole literature.
Specifically, SnO$_2$ is deposited onto the substrate before the perovskite layer based on the literature context.
By grounding the extraction process in this literal knowledge of device physics (e.g., the $n-i-p$ framework), our pipeline ensures the semantic integrity of the research, providing a reliable foundation for subsequent autonomous optimization.

\subsubsection{Analysis on Simulator Tasks}

\noindent$\bullet$ \textbf{Experiment Setting}.
To visualize the high-dimensional parameter space navigated by PVK-BO, we conduct a statistical correlation analysis on the simulator data (Fig.~\ref{fig:ablation_comparison_AB} (e)).
This analysis serves as a foundation for understanding the landscape of material properties optimization.
The analysis covers two experiments in the Sec.~\ref{sec:performance_of_simulation_experiment}.
First, Band Alignment (left panel) focuses on the energy band position matching of the PVK and carrier transport layers, involving five key variables: the electron affinity of the perovskite absorber ($\chi_{PVK}$), and the tunable energy parameters of the transport layers, specifically the electron affinities ($\chi_{HTL}, \chi_{ETL}$) and bandgaps ($E_g^{HTL}, E_g^{ETL}$).
Second, Doping optimization and interface defect optimization (right panel) focuses on interface carrier transport and recombination dynamics, involving eight variables: the interface defect densities at both the front and rear of PVK layer interfaces ($N_t^{PVK/ETL}, N_t^{HTL/PVK}$), as well as the acceptor/hole ($N_a$) and donor/electron ($N_d$) concentrations of the PVK, HTL, and ETL layers, respectively ($N_{a/d}^{PVK}, N_{a/d}^{HTL}, N_{a/d}^{ETL}$).
To elucidate the characteristics of optimal recipes, we categorize the simulator outcomes into two distinct clusters based on Power Conversion Efficiency (PCE): High-Performance samples (PCE $> 20\%$, marked in red) representing optimization targets, and Low-Performance samples (PCE $\leq 20\%$, marked in blue) representing suboptimal regions.

\noindent$\bullet$ \textbf{Results}.
The distribution differences between the red and blue clusters reveal how PVK-LLM leverages domain knowledge to identify sweet spots.
In the Band Alignment task, the scatter plot matrix reveals a strong coupling effect.
The red dots (high PCE) gather in distinct areas instead of being scattered randomly.
Specifically, in the $\chi_{HTL}$ vs. $\chi_{ETL}$ subplot, devices are confined to a specific range, suggesting that these two parameters must be tuned synergistically.
This indicates that parameters must be tuned synergistically to achieve precise interfacial matching rather than optimized individually.
In the Doping Optimization task, the analysis follows specific device semiconductor physics laws.
For instance, within the ETL, we observe a strong positive correlation between performance and donor (electron) concentration ($N_d^{ETL}$), contrasted with a negative correlation for acceptor (hole) concentration ($N_a^{ETL}$).
This aligns with the theoretical requirement that high n-type doping is crucial for electron selectivity and conductivity.
From an algorithmic perspective, the contrast between the broad, scattered distribution of blue dots (exploration) and the highly localized, high-yield islands of red dots (exploitation) demonstrates the efficiency of PVK-BO.
By assimilating physical experimental results from the search trajectory, the framework avoids wasting iterations on low-performance regions and converges onto the narrow physical property windows where high efficiency is attainable.

\subsection{Analysis on PVK-BO}

To investigate the PVK-BO framework's efficiency, we conduct an ablation study focusing on the initialization, surrogate model efficacy, and prompt-based context, as illustrated in Fig.~\ref{fig:ablation_BO}. 

\subsubsection{Analysis on Initialization}

\noindent$\bullet$ \textbf{Experiment Setting}.
To evaluate the efficacy of the initialization phase, we conduct a comparative analysis between PVK-LLM and three space-filling strategies: Random search, Sobol sequences, and Latin Hypercube (LH) sampling (Fig.~\ref{fig:ablation_BO} (a)). 
This experiment aims to quantify the model's ability to alleviate the cold-start problem inherent in high-dimensional optimization. 
We employ two key metrics to assess the quality of the initial candidates.  
First, we define Initial Best Regret as the difference between the global optimum and the best value found in the initial batch. 
This metric evaluates the optimality gap, reflecting how close the model's best initial guess is to the theoretical limit.
Second, we calculate Hit Rate as the percentage of samples exceeding a high-performance threshold (PCE $> 20\%$). 
This metric evaluates the exploration robustness, quantifying the density of valid, high-quality candidates identified within the sparse high-dimensional search space.

\noindent$\bullet$ \textbf{Results}.
We experiment with five independent runs.
It is noteworthy that Random initialization outperforms structured space-filling designs (Sobol and Latin Hypercube) in both initial best regret and hit rate. 
This phenomenon can be attributed to the low effective dimensionality inherent in perovskite recipe optimization~\cite{bergstra2012random}. 
While Sobol and LH prioritize uniform coverage across the entire high-dimensional hypercube, their geometric structure may struggle to capture the critical features distributed irregularly within low-dimensional subspaces. 
In contrast, Random search, despite its simplicity, avoids the rigid structural constraints of quasi-random sequences, allowing for more flexible exploration of the rugged and discontinuous material landscape. 
However, both blind search strategies lag behind PVK-LLM, which leverages domain knowledge to directly target promising regions.
PVK-LLM achieves the lowest Initial Best Regret ($4.24$) and the highest Hit Rate ($13.3 \%$). 
It illustrates the advantage of PVK-LLM in reducing exploration blindness, proving that it can leverage its built-in domain knowledge to lock onto valid, high-performance recipes at the very start of the search space, thereby overcoming the cold-start problem prevalent in standard optimization algorithms~\cite{van2024traversing}.

\subsubsection{Analysis on Surrogate Model}
\label{sec:analysis_on_surrogate_model}

\begin{figure*}
    \centering
    \includegraphics[width=1.0\linewidth]{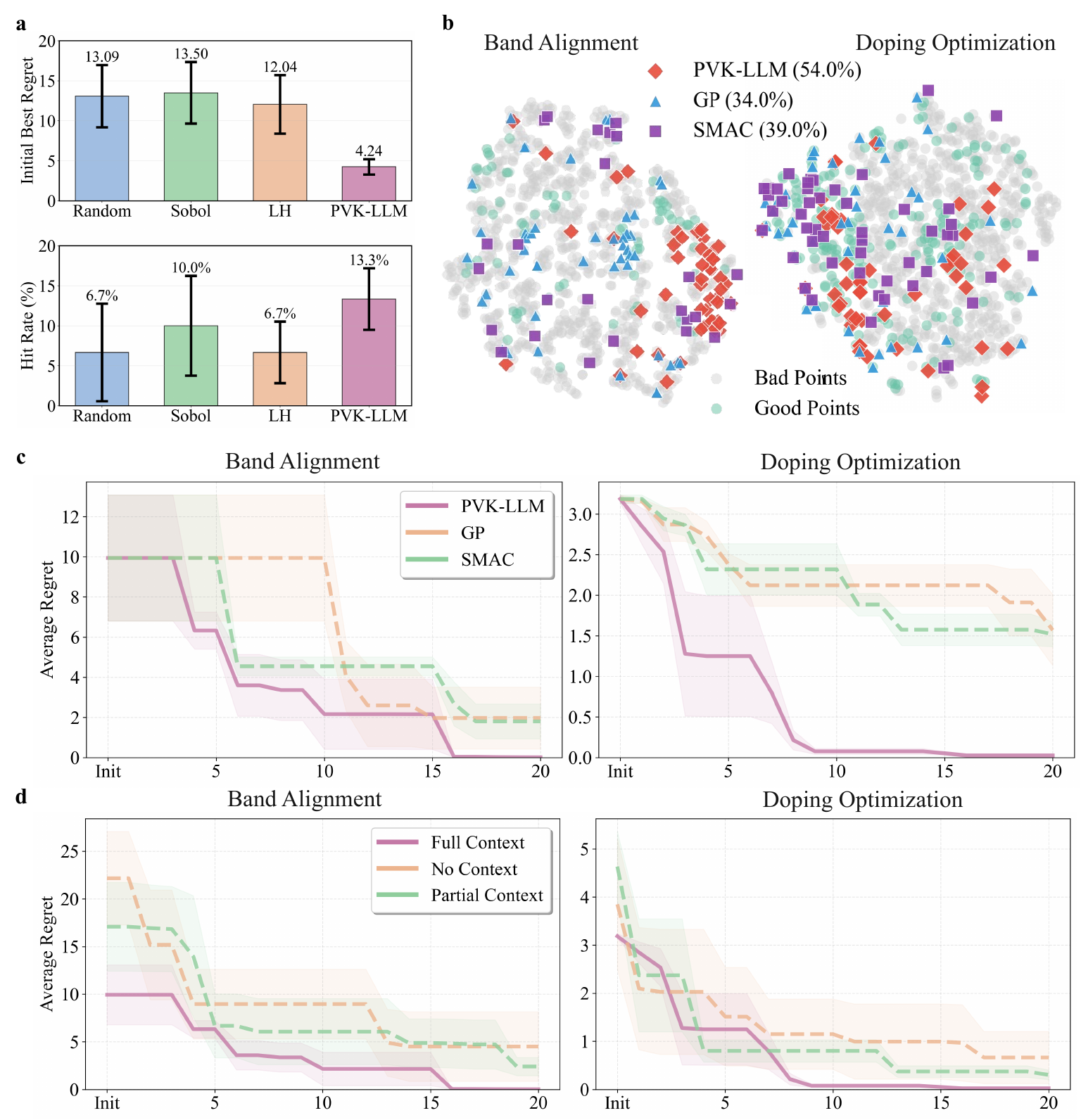}
    \caption{Robustness analysis of the optimization strategy. 
    All experiments are conducted with five independent runs.
    (a) Comparison of Initial Best Regret and Hit Rate. PVK-LLM achieves the lowest initial regret ($4.24$) and the highest hit rate ($13.3 \%$), indicating its ability to recommend valid high-performance recipes from the start.
    The Hit Rate defined as the proportion of initial recommendations that achieve a Power Conversion Efficiency (PCE) exceeding $20 \%$.
    LH denotes the Latin Hypercube.
    (b) Visualization of the search space coverage. 
    PVK-BO explores high-potential regions more effectively compared to Gaussian Process (GP) and SMAC baselines. 
    (c) Average Regret curves for Band Alignment and Defects Doping tasks. 
    The decline in regret for PVK-LLM confirms its accelerated convergence speed compared to traditional methods. 
    (d) Impact of context on optimization. 
    Comparing full context, partial context, and no context settings reveals that providing the model with experimental history and domain context lowers regret and improves optimization outcomes, supporting the use of In-Context Learning (ICL). }
    \label{fig:ablation_BO}
\end{figure*}

\noindent$\bullet$ \textbf{Experiment Setting}.
To evaluate the efficiency of the surrogate model, we conduct a comparative assessment between the PVK-LLM-based surrogate and traditional Bayesian Optimization baselines, specifically Gaussian Process (GP) and SMAC~\cite{lindauer2022smac3}.
The performance is analyzed through Fig.~\ref{fig:ablation_BO}: the search space coverage distribution (Fig.~\ref{fig:ablation_BO} (b)) and the optimization convergence trajectories (Fig.~\ref{fig:ablation_BO} (c)).
We employ two quantitative metrics to measure efficiency:
(1) High-Potential Exploration Rate. 
We define it as the percentage of sampling points falling within the ``Good Points" clusters (characterized by PCE $> 20\%$). 
This metric evaluates the surrogate model's ability to direct the search towards high-yield regions.
(2) Average Regret. Its tracks the ``distance to perfection.'' 
It represents the gap between the theoretical best possible efficiency (global optimum) and the best result the model has found so far.
We average this gap over multiple independent experiments to ensure reliability. 
A drop in this value means the model learns quickly and gets closer to the optimal solution with fewer wasted steps.

\noindent$\bullet$ \textbf{Results}.
The comparative results demonstrate the superior intelligence of the PVK-LLM surrogate.
As visualized in the search space embedding (Fig.~\ref{fig:ablation_BO} (b)), PVK-LLM's sampling points are densely concentrated within the high-performance regions.
Quantitatively, it achieves a High-Potential Exploration Rate of $54.0\%$, outperforming both GP ($34.0\%$) and SMAC ($39.0\%$).
This precision indicates that PVK-LLM utilizes internalized domain knowledge to construct a more accurate prior of the material landscape, thereby avoiding low-utility regions.
This advantage translates into superior convergence efficiency, as shown in the regret curves (Fig.~\ref{fig:ablation_BO} (c)).
In both Band Alignment and Doping Optimization tasks, PVK-LLM exhibits the fastest decline rate and the lowest final regret.
This confirms that by dynamically assimilating physical experimental results from experimental feedback, the model can navigate high-dimensional spaces with complex physical constraints more efficiently than traditional probabilistic models.

\subsubsection{Analysis on Prompt Engineering}

\noindent$\bullet$ \textbf{Experiment Setting}.
To investigate the impact of context quality on the optimization process, we design an ablation study involving three distinct prompt modes, as shown in Fig.~\ref{fig:ablation_BO} (d).
The Full Context mode represents the complete integration of the prompt in PVK-BO, where the model is provided with both task-specific physical descriptions (e.g., device architecture) and historical experimental feedback.
In contrast, the Partial Context mode provides only the physical meta-features of the task, and the No Context mode offers only a generic optimization objective without specific constraints. 
The common limitation shared by both the No Context and Partial Context modes is the absence of historical experimental records. 
This means that regardless of whether the physical task background is provided, both modes suffer from a deficiency in physical experiment results. 
To quantify the impact of these distinct knowledge levels on the optimization trajectory, we employ the Average Regret metric. 
As established in Sec.~\ref{sec:analysis_on_surrogate_model}, this metric tracks the optimization gap over iterations, where a lower and faster-converging curve indicates a more effective utilization of context to guide the search process.

\noindent$\bullet$ \textbf{Results}.
The results reveal the role of these information components. 
It is important to note that all three modes utilize the same PVK-LLM, and thus all possess the underlying domain knowledge. 
However, in the No Context mode, the lack of task-specific descriptors prevents the model from aligning its theoretical knowledge with the specific experimental goals, causing it to degrade into a generic search tool. 
In the Partial Context mode, while the physical background activates relevant theoretical priors to ensure valid recipe generation, the specific absence of physical experiment feedback prevents the model from learning refinement strategies from experience. 
Consequently, without the guidance of historical data, the model cannot perform ``exploitation" effectively, leading to a marked increase in regret compared to the Full Context mode. 
This analysis supports that while domain knowledge (activated by physical context) is essential for a valid warmstarting, physical experiment feedback (derived from historical feedback) is important for the precise optimization path.

\section{Discussion}
In this study, we propose PVK-LLM, a domain-knowledge-guided Large Language Model capable of bridging the semantic gap between natural language, scientific literature, and materials recipes and experimental parameters in perovskite research.
To address the limitation that general LLMs often lack the material-recipe and experimental-parameter knowledge required for the PSC material science, we employ a three-stage curricular learning strategy to inject \textbf{domain knowledge}, transforming the general LLM into a specialized domain expert.
By integrating the Bayesian Optimization engine (PVK-BO), PVK-LLM transforms the experimental records that directly guide subsequent experimental iterations.
Our experiments demonstrate the effectiveness of this framework across the benchmark and simulator experiments.
PVK-LLM is not only capable of understanding complex fabrication parameters but also serves as a proactive research copilot, guiding the advancement of high-performance devices.
Specifically in wet-lab experiments, our framework, guided by domain knowledge, identifies the processing bottleneck and navigates a high-dimensional parameter space to achieve a \textbf{PCE of $26.00\%$}.
These results indicate our framework's potential in material recipe design, offering valuable insights to researchers and accelerating the iteration cycle from theoretical design to laboratory validation.

Despite these advancements, two limitations remain.
One limitation lies in the constraints imposed by the quality and availability of open-source academic literature.
While PVK-LLM captures domain knowledge, the model's predictive accuracy is bounded by the data noise and reporting inconsistencies inherent in the existing publications.
Another limitation is that the physical execution currently relies on a human-in-the-loop workflow, although our framework achieves a closed loop in terms of decision-making.
This manual intervention limits the throughput compared to fully automated robotic platforms~\cite{liu2025balancing,zhao2025embedding}.
Based on the current architecture, PVK-LLM exhibits strong extensibility.
By incorporating standard API interfaces for laboratory automation equipment, the framework can be adapted to drive self-driving laboratories, thereby enabling continuous autonomous experimentation~\cite{sanders2023biological,portner2025actor}.

Finally, the versatility of our domain-knowledge-guided framework extends beyond the specific application of perovskite solar cells. 
The core methodology synergizes domain knowledge with experiment results and applies to other high-dimensional material challenges, such as battery electrolyte optimization or organic photovoltaic light-active materials screening. 
Furthermore, by incorporating interfaces for laboratory automation, the framework can be designed to evolve into a foundational agent for self-driving laboratories.
The enrichment of training data and the integration of robotic execution with high accuracy for experiments and better repeatability would enhance the capability of PVK-LLM. 
The enhanced PVK-LLM can reverse guide and facilitate continuous autonomous experimentation. 
This trajectory not only accelerates material innovation but also lays the groundwork for autonomous scientific discovery in the physical sciences.

\section{Methods}

\subsection{Corpus Collection}
\label{sec:corpus_collection}

% 收集了哪些文章
We collect the corpus of $4,726$ scientific articles published in the leading journals. 
These publications are obtained through an automated pipeline built on an API.
The search is constrained to full-length and review articles published since $2018$, utilizing the keywords ``perovskite",  ``perovskite solar cell(s)", and ``PVK”. 
For each retrieved publication, bibliographic metadata (including titles, authors, DOIs, journal names, and publication dates) is extracted, while the full texts are converted to a Markdown format. 

% % 我们通过文章,制作了三块数据集,分别是简单问答数据集,指令跟随数据集,KG
% \noindent \textbf{Dataset Construction.} 
% We construct three part datasets based on the corpus for the different training stage, including Foundational QA Dataset (PVK-Sci), Task Instruction Dataset (PVK-Instruct) and Knowledge Graph (PVK-KG).
% % 简单问答数据集怎么制作的
% % 让研究人员设计7 research categories, 21 research questions
% % 针对这些常见问题，使用智能体进行问题抽取，并且使用智能体进行问题总结和问题纠正
% PVK-Sci contains $55,101$ instances around $4.4$ million tokens.
% % 指令跟随数据集怎么制造的
% % 旨在包含两部分能力，一部分是长引用
% PVK-Instruct.

% KG 怎么制作的

\subsection{Curriculum Learning Stage}

% 三阶段训练
The training of PVK-LLM follows a systematic framework inspired by the principles of curriculum learning~\cite{wang2021survey}, utilizing Qwen2.5-32B~\cite{qwen2.5} as the backbone. 
This methodology is designed to incrementally build the model's capabilities, progressing from foundational domain knowledge to complex analytical skills, and finally to dynamic, verifiable reasoning. 
The curriculum is structured into three distinct stages: Domain Knowledge Injection, Capability-Instruction Alignment, and Knowledge Retrieval Augmentation.
To align with this strategy, we construct specialized datasets corresponding to each stage: the Foundational QA Dataset (PVK-Sci), the Task Instruction Datasets (PVK-Cite and PVK-Exp), and the Knowledge Graph (PVK-KG).

\subsubsection{Domain Knowledge Injection}

\noindent $\bullet$ \textbf{Dataset}. 
For constructing PVK-Sci, we collaborated with five domain researchers to design a comprehensive taxonomy for perovskite research.
As illustrated in Fig.~\ref{fig:supp_1} (b), we identified $7$ thematic research categories (e.g., ``Structure \& Fab'', ``Stability Improve.'', ``Defect \& Recomb.'') which encompass $21$ specific research questions (Q1--Q21).
These questions cover topics ranging from precursor preparation (Q2) to interface wettability (Q6) and intrinsic stability mechanisms (Q13--Q15).
The corpus from Sec.~\ref{sec:corpus_collection} is processed by a three-phase GPT-4o-based system for high-quality data synthesis: an Information Extraction Agent pulls relevant data chunks, a Quality Validation Agent filters for accuracy, and a Document Summarization Agent synthesizes the facts into coherent answers.
The prompt in the construction pipeline is shown in Fig.~\ref{fig:prompt} (a).
This process is repeated for all $21$ research questions to construct the complete PVK-Sci dataset.
The token length distribution analysis of the prompts and responses in PVK-Sci is presented in Fig.~\ref{fig:combined_length_distribution}, which guides our maximum sequence length setting.

\noindent $\bullet$ \textbf{Training Strategy}.
The initial stage focuses on injecting domain knowledge into the base model to establish a foundational understanding of perovskite science. 
As illustrated in Fig.~\ref{fig:overview} (b), the model utilizes a stacked Transformer decoder architecture, incorporating masked self-attention and feed-forward networks (FFN) with residual connections and layer normalization.
Fine-tuning on PVK-Sci ($\mathcal{D}_{sci}$) ensures the model internalizes the core vocabulary and semantic patterns of the domain.
The instruction tuning is optimized under a causal language modelling objective:
\begin{equation}
\label{equ:training_strategy}
\begin{aligned}
\scalebox{0.92}{$
\min_{\theta} \mathbb{E}_{(x, y) \sim \mathcal{D}} [\mathcal{L}_{CE}(LM_{\theta_{model}}(x), y)],
$}
\end{aligned}
\end{equation}
where $x$ and $y$ denote the input and ground truth (GT), respectively, sampled from the dataset $\mathcal{D} = \mathcal{D}_{sci}$.

\subsubsection{Capability-Instruction Alignment}

\noindent $\bullet$ \textbf{Dataset}.
To equip the model with comprehensive research capabilities, we construct two complementary instruction datasets:
\begin{enumerate}
    \item \textbf{PVK-Cite (Perovskite Knowledge Citation dataset)}: This dataset $\mathcal{D}_{cite}$ focuses on long-context document analysis, enabling the model to generate responses with fine-grained, sentence-level citations~\cite{zhang2025longcite}. We leverage core domain keywords (e.g., ``PCE'', ``stability'') to retrieve knowledge tuples from PVK-KG and assemble them into long documents.
    Using a ``coarse-to-fine'' annotation principle, GPT-4o generates answers, first identifying chunk-level support and then pinpointing specific sentence-level sources.
    \item \textbf{PVK-Exp (Perovskite Experiment Analysis dataset)}: This dataset $\mathcal{D}_{exp}$ targets the interpretation of scientific tables. 
    We collect a diverse set of experimental tables from the literature. 
    These tables are paired with instruction-following data that requires the model to extract key performance metrics (e.g., PCE values) and analyze physical trends.
\end{enumerate}

\noindent $\bullet$ \textbf{Training Strategy}.
The second stage fine-tunes PVK-LLM on both \textbf{PVK-Cite} and \textbf{PVK-Exp}. 
This fine-tuning empowers the model to not only ground its textual answers in literature but also directly analyze raw experimental data.
The instruction tuning is optimized under the same causal language modelling objective as Eq.~\ref{equ:training_strategy}, where $\mathcal{D} = \mathcal{D}_{cite} \cup \mathcal{D}_{exp}$.

\subsubsection{Knowledge Retrieval-Augmented Generation}
\label{sec:knowledge_retrieval_augmentation}

\noindent $\bullet$ \textbf{Dataset}.
Drawing upon expert knowledge, we develop the PVK-KG schema to structure unstructured literature.
As shown in Fig.~\ref{fig:supp_1} (a), the schema is centered around core entities including ``Fabrication'', ``Performance'', and ``Parameters''.
These entities are defined by specific attributes and data types; for instance, ``Performance'' aggregates metrics such as ``Voc Value'' (Float), ``PCE Value'' (Float), and ``Light Stability'' (String).
Similarly, ``Fabrication'' links to ``Method'' (e.g., spin coating) and ``Annealing Parameter''.
As shown in Fig~\ref{fig:kg_supp2} (a), we extract sub-ontologies based on this schema and employ GPT-4o to query document collections, extracting domain knowledge into a structured graph. 
Subsequently, we construct quality control through entity disambiguation and relationship deduplication to resolve semantic ambiguities and merge redundant entries. 
The pipeline concludes by instantiating the knowledge graph within a graph database, incorporating citation relationships to enhance data provenance and knowledge grounding.
We compare PVK-KG with other chemistry KG or Material KG~\cite{mrdjenovich2020propnet,jacobsson2022open,statt2023materials,venugopal2024matkg,matportal2024,nie2022automating,zhang2024materials} in Fig.~\ref{fig:kg_supp2} (b).
Unlike general material knowledge graphs such as MatKG~\cite{venugopal2024matkg} or MKG that primarily index broad concepts, our system targets specific device parameters defined by the schema. 
Furthermore, in contrast to databases like Perovskite DB that achieve similar granularity through likely manual curation, PVK-KG leverages Large Language Models for automated information extraction from unstructured scientific papers. 
This approach enables the efficient processing of extensive literature to generate $23,789$ entities and $22,272$ triples while maintaining the depth of data required for analyzing device performance.

\noindent $\bullet$ \textbf{Inference Strategy}.
In the third stage, Knowledge Retrieval-Augmentated Generation (RAG) connects PVK-LLM to PVK-KG, providing dynamic access to the latest research to resolve knowledge conflicts and static limitations.
To ensure comprehensive retrieval, our framework adopts a Dual-Level Retrieval paradigm inspired by LightRAG~\cite{guo2024lightrag}.
The process begins with query decomposition to extract two types of keywords, including low-level and high-level keywords.
In low-level keywords, specific entities (e.g., material names like ``MAPbI$_{3}$'' or specific metrics) facilitate local graph traversal to acquire directly related neighboring nodes.
In high-level keywords, broader topics or intents (e.g., ``interfacial recombination mechanism'') guide the retrieval of high-order relatedness and broader conceptual neighborhoods.
By combining the graph fragments retrieved from both low-level and high-level keywords, we construct a comprehensive context. 
This context is re-ranked and fed into PVK-LLM, ensuring the generated answers are both detail-oriented and structurally holistic.

% 贝叶斯优化
% 钙钛矿太阳能电池不稳定，复现困难，常常因为环境问题（例如湿度，温度）的不确定，而导致很多文献的参数实现困难，所以我们提出了一套因地制宜的参数优化策略。先给出粗参数再选择那个参数进行优化，最后使用贝叶斯优化对参数进行详细优化
\subsection{Optimization Framework}
\label{sec:optimization_application}
PSCs involve a high-dimensional material recipe space, making exhaustive exploration experimentally intractable. 
To address this, we propose a hierarchical framework adaptable to two distinct experimental settings:
\begin{enumerate}
    \item \textbf{Wet-Lab Workflow:} A three-stage active learning pipeline that progressively narrows the search scope from broad exploration to focused optimization.
    \item \textbf{Simulator Workflow:} A targeted optimization process driven directly by the specialized PVK-BO engine.
\end{enumerate}
Below, we detail the three-stage workflow designed for wet-lab challenges, followed by the specific mechanics of the core PVK-BO engine used in both settings.

\subsubsection{Workflow for Wet-Lab Experiments}
As illustrated in Fig.~\ref{fig:wet-lab} (a), to effectively navigate the high-dimensional parameter space, we design a three-step active learning workflow~\cite{ji2023self}. 
This process progressively focuses the search scope from a broad baseline to specific critical parameters.

\noindent $\bullet$ \textbf{Step I: Initialization}. 
The process begins with the cold-start phase, where no project-specific data exists. 
Instead of random guessing, the PVK-LLM leverages its pre-trained domain knowledge to generate a feasible baseline recipe based on the task description $\mathcal{T}$. 
For instance, as shown in Fig.~\ref{fig:wet-lab} (a), the model suggests a specific precursor solution (e.g., mixing 3MTPAI and PDAI$_{2}$) tailored to the user's specific request for a single-junction normal bandgap device.

\noindent $\bullet$ \textbf{Step II: Screening}. 
Once the baseline recipe is tested, the experimental feedback is fed back to the LLM. 
The model performs a sensitivity analysis to identify which parameters are most critical for performance improvement in the current context~\cite{grosjean2025network}. 
As depicted in Fig.~\ref{fig:wet-lab} (a), the model reasons that interface passivation is the bottleneck and identifies specific molecules (e.g., 3MTPAI, PDAI$_2$, EDAI$_2$, and PipDI) as the critical variables to optimize, effectively reducing the optimization dimensionality.
We define these critical variables as the optimization subspace $\mathcal{S}$.

\noindent $\bullet$ \textbf{Step III: Optimization}. 
Targeting the identified optimization subspace $\mathcal{S}$, we enter the automated optimization phase. 
Here, we deploy the specific PVK-BO algorithm to iteratively explore the reduced search space and locate the global optimum.

\subsubsection{PVK-BO Algorithm}
\label{sec:pvk_bo_algorithm}
The core engine driving both Step III of the wet-lab workflow and the simulator experiments is our proposed PVK-BO.
As illustrated in the closed-loop workflow of Fig.~\ref{fig:sim_exp} (a), this algorithm employs a fully LLM-driven Bayesian Optimization cycle. 
Once the subspace $\mathcal{S}$ is determined, the algorithm executes with four system inputs: the subspace $\mathcal{S}$, the task description $\mathcal{T}$, the optimization target $y_{target}$, and the optimization step $t_{limit}$.
The prompt structure used in the pipeline is shown in Fig.~\ref{fig:prompt} (c).
We formulate the six key steps as follows:

\begin{enumerate}
    \item \textbf{Warmstarting}. 
    To mitigate the cold-start problem, we initialize the optimization within the subspace $\mathcal{S}$.
    PVK-LLM generates the first batch of recipes $\mathbf{x}_0$ by conditioning on the task $\mathcal{T}$ and the subspace constraints $\mathcal{S}$:
    \begin{equation}
    \mathbf{x}_0 \sim \text{LLM}(\mathcal{S}, \mathcal{T}).
    \end{equation}

    \item \textbf{Experiment Evaluation}. 
    The synthesized recipe (either $\mathbf{x}_0$ or $\mathbf{x}_{t+1}$) is physically characterized to obtain the ground-truth performance $y_{t+1}$. 
    The history is updated, and the iteration counter advances ($t \leftarrow t+1$):
    \begin{equation}
    \begin{aligned}
    y_{t+1} &\sim \text{Experiment}(\mathbf{x}_{t+1}) \\
    \mathcal{H}_{t+1} &\leftarrow \mathcal{H}_t \cup \{(\mathbf{x}_{t+1}, y_{t+1})\}.
    \end{aligned}
    \end{equation}
    
    \item \textbf{Candidate Sampling}. 
    In this phase, PVK-LLM functions as a generative proposal distribution. 
    By conditioning on the desired high-performance target $y_{target}$ and the observed history $\mathcal{H}_t$, we leverage In-Context Learning to perform inverse design~\cite{yao2021inverse}:
    \begin{equation}
    \mathcal{C}_t = \{ \tilde{\mathbf{x}}_t \mid \tilde{\mathbf{x}}_t \sim \text{LLM}(y_{target}, \mathcal{H}_t) \}.
    \end{equation}
    This step generates a batch of candidate recipes $\mathcal{C}_t$ that are semantically aligned with the target performance.

    \item \textbf{Surrogate Modelling}. 
    The LLM acts as a probabilistic surrogate to predict the expected performance $\hat{y}_t$ for each candidate in $\mathcal{C}_t$ before physical verification~\cite{kulichenko2023uncertainty}:
    \begin{equation}
    \hat{y}_t \sim \text{LLM}(\tilde{\mathbf{x}}_t, \mathcal{H}_t).
    \end{equation}
    Here, $\hat{y}_t$ represents the predicted efficiency for a candidate $\tilde{\mathbf{x}}_t$, incorporating the model's uncertainty.

    \item \textbf{Acquisition Function}. 
    Instead of relying on rigid statistical functions (e.g., Expected Improvement), we deploy PVK-LLM itself as an intelligent acquisition function. 
    To balance exploration and exploitation, PVK-LLM evaluates the utility $\alpha_{\text{LLM}}$ of each candidate pair $(\tilde{\mathbf{x}}_t, \hat{y}_t)$ by weighing the potential performance gain against the exploration value~\cite{sabanza2025best}. 
    The optimal recipe for the next iteration is selected by:
    \begin{equation}
    \mathbf{x}_{t+1} = \operatorname*{arg\,max}_{\tilde{\mathbf{x}}_t \in \mathcal{C}_t} \alpha_{\text{LLM}}( \tilde{\mathbf{x}}_t, \hat{y}_t).
    \end{equation}
    This allows the algorithm to leverage domain knowledge to select the most promising recipe.
    The selected recipe $\mathbf{x}_{t+1}$ is then forwarded to the \textbf{Experiment Evaluation} module for physical verification, thereby closing the optimization loop.

    \item \textbf{Exit}. 
    The loop terminates if the target is met ($y_{t+1} > y_{target}$) or the optimization step is exhausted ($t > t_{limit}$).
\end{enumerate}

\section*{Data Availability}

All datasets used in this paper are publicly available via the original publications and releases. Source datasets are also provided at: \url{https://sites.google.com/view/pvk-llm/dataset}.

\section*{Code Availability}

Our code was implemented using the Python (v3.10.15), PyTorch (v2.5.1), and Qwen2.5-32B for LLM components. 
Our codes are available at: \url{https://sites.google.com/view/pvk-llm/code}.

\section*{Acknowledgments}
This work is supported by the Advanced Materials-National Science and Technology Major Project (Grant No. 2025ZD0620100).

\section*{Author Contributions}
P.S. and C.Y. conceived the study and supervised the project. 
P.S. implemented the PVK-LLM and PVK-BO.  
S.C. designed the wet-lab experiments. 
L.Z. designed the simulator experiments.
X.L. and H.Y. helped collect the data. 
Y.Z., X.C. and T.Z. contributed to the analysis of results and provided valuable insights.
P.S. wrote the manuscript with inputs and guidance from all the authors.

\section*{Competing Interests}

The authors declare no competing interests.

\clearpage


\begin{appendices}

\section{Details of Dataset Curation}\label{app:dataset_curation}

\subsection{Schema on PVK-Sci and PVK-KG}
\begin{figure*}
    \centering
    \includegraphics[width=1.0\linewidth]{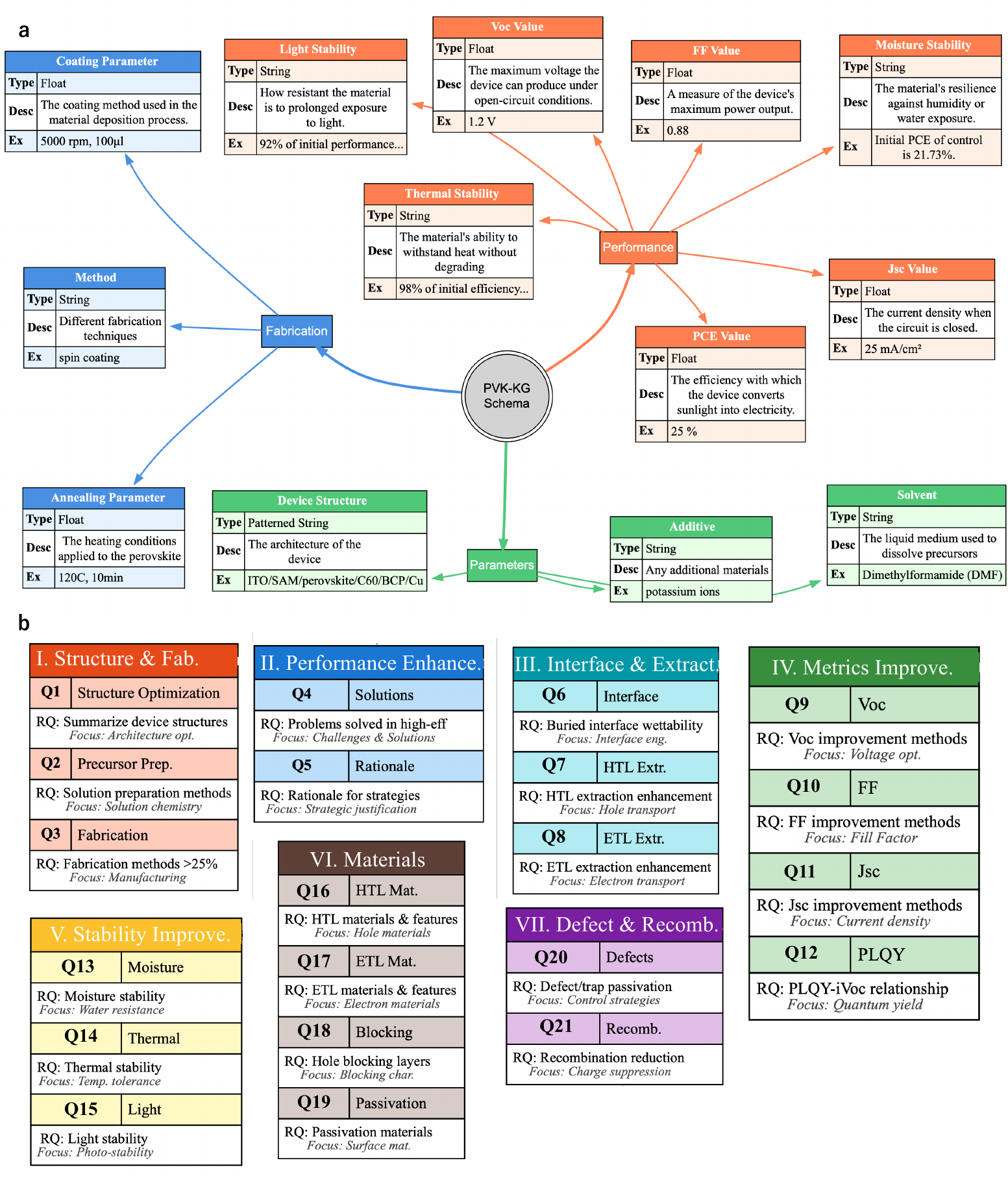}
    \caption{Schema definition for the PVK-KG and research question taxonomy for the PVK-Sci. (a) The data schema of the PVK-KG, illustrating the hierarchical relationships between core entities (Fabrication, Performance, and Parameters) and their specific attributes. 
    (b) Systematic classification of research questions (RQ) in PVK-Sci. The $21$ identified research questions (Q1–Q21) are categorized into seven thematic domains, ranging from structure and fabrication optimization (I) to defect passivation and recombination suppression (VII).
    The following abbreviations are used in this figure: Fabrication (Fab.), Enhancement (Enhance.), Extraction (Extract.), and Materials (Mat.).
    }
    \label{fig:supp_1}
\end{figure*}

In Fig.~\ref{fig:supp_1}, we visualize the foundational architecture for the PVK-LLM framework, establishing the structural and categorical rules required for precise scientific knowledge extraction. 
It is divided into two sections: a detailed data schema for the PVK-KG and a systematic taxonomy of research questions for the PVK-Sci dataset.
In Fig.~\ref{fig:supp_1} (a), the hierarchical data schema for PVK-KG is presented, organized around three core ontologies: Fabrication, Performance, and Parameters. 
Each entity is meticulously defined with specific attributes and data types to maintain high granularity for machine interpretation; for example, ``Performance" aggregates metrics like Voc Value, PCE Value (Float), and Moisture Stability (String), while ``Fabrication" connects to experimental variables such as Coating Parameters and Annealing conditions. 
This schema allows an automated pipeline to convert unstructured scientific literature into a structured graph.
Fig.~\ref{fig:supp_1} (b) establishes the PVK-Sci research question taxonomy, which is critical for the ``Knowledge Injection" stage of the model’s training. 
The taxonomy systematically classifies $21$ distinct research questions (Q1–Q21) into seven thematic domains, ranging from ``Structure \& Fabrication" and ``Materials" to ``Defect \& Recombination". 
These predefined questions guide the synthesis of high-quality data chunks from the literature corpus, covering topics such as precursor preparation, interface wettability, and intrinsic stability mechanisms.

\subsection{PVK-KG Construction}
\begin{figure*}[htbp]
  \centering
  
  % 1. 插入图片
  \includegraphics[width=1.0\textwidth]{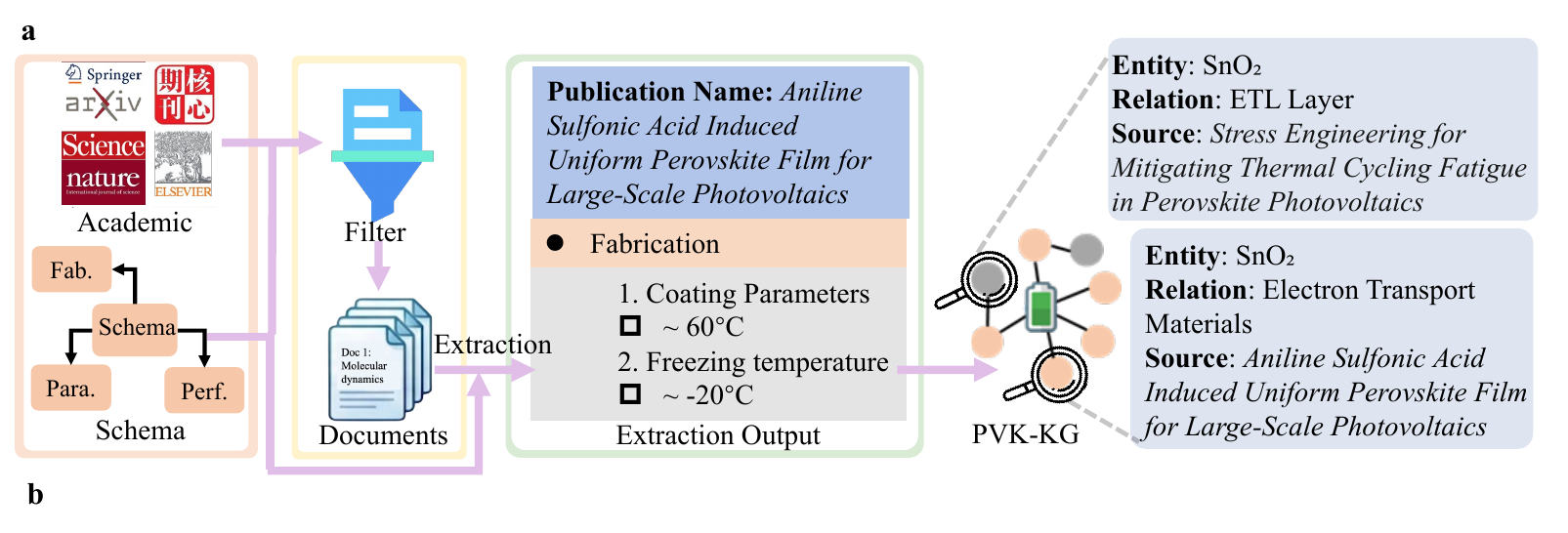}
  
  \vspace{0.5em} % 增加垂直间距

  % 2. 插入表格
  \resizebox{1.0\textwidth}{!}{ 
        \begin{tabular}{@{}c|c|c|c|c|c|c|c@{}}
\toprule
\textbf{Name}                    & \textbf{Venue}                                                & \textbf{Year}                  & \textbf{Scope}                                              & \textbf{Scale}                                                                                     & \textbf{Source}                  & \textbf{Automatic}            & \textbf{Granularity}                                                                   \\ \midrule
Propnet~\cite{mrdjenovich2020propnet}                          & Matter                                                        & 2020                           & Materials                                                   & \begin{tabular}[c]{@{}c@{}}100+   \\ Modules\end{tabular}                                          & Database                         &                               & Concepts                                                                               \\ \midrule
Li-ion KG~\cite{nie2022automating}                        & \begin{tabular}[c]{@{}c@{}}Adv. Funct. \\ Mater.\end{tabular} & 2022                           & \begin{tabular}[c]{@{}c@{}}Li-ion \\ Batteries\end{tabular} & \begin{tabular}[c]{@{}c@{}}20M\\ Triples\end{tabular}                                              & Papers                           &                               & Concepts                                                                               \\ \midrule
Perovskite DB~\cite{jacobsson2022open}                    & \begin{tabular}[c]{@{}c@{}}Nature \\ Energy\end{tabular}      & 2022                           & Perovskite                                                  & \begin{tabular}[c]{@{}c@{}}42k\\ Devices\end{tabular}                                              & Papers                           &                               & \begin{tabular}[c]{@{}c@{}}Device\\ Parameters\end{tabular}                            \\ \midrule
MEKG~\cite{statt2023materials}                             & \begin{tabular}[c]{@{}c@{}}Digital \\ Discovery\end{tabular}  & 2023                           & Materials                                                   & \begin{tabular}[c]{@{}c@{}}52M\\ Nodes\end{tabular}                                                & Experiments                      &                               & \begin{tabular}[c]{@{}c@{}}Data\\ Logs\end{tabular}                                    \\ \midrule
MKG~\cite{zhang2024materials}                              & \begin{tabular}[c]{@{}c@{}}Scientific \\ Data\end{tabular}    & 2024                           & Materials                                                   & \begin{tabular}[c]{@{}c@{}}8k\\ Entites\end{tabular}                                               & Papers                           & Yes                           & Concepts                                                                               \\ \midrule
MatKG~\cite{venugopal2024matkg}                            & \begin{tabular}[c]{@{}c@{}}Scientific \\ Data\end{tabular}    & 2024                           & Materials                                                   & \begin{tabular}[c]{@{}c@{}}70k\\ Nodes\end{tabular}                                                & Papers                           & Yes                           & Concepts                                                                               \\ \midrule
TFSCO~\cite{matportal2024}                            & MatPortal                                                     & 2024                           & \begin{tabular}[c]{@{}c@{}}Solar\\ Cells\end{tabular}       & \begin{tabular}[c]{@{}c@{}}651\\ Classes\end{tabular}                                              & Database                         &                               & Concepts                                                                               \\ \midrule
\multirow{2}{*}{\textbf{PVK-KG}} & \multirow{2}{*}{\textbf{(Ours)}}                              & \multirow{2}{*}{\textbf{2025}} & \multirow{2}{*}{\textbf{Perovskite}}                        & \multirow{2}{*}{\textbf{\begin{tabular}[c]{@{}c@{}}23,789 Entities\\ 22,272 Triples\end{tabular}}} & \multirow{2}{*}{\textbf{Papers}} & \multirow{2}{*}{\textbf{Yes}} & \multirow{2}{*}{\textbf{\begin{tabular}[c]{@{}c@{}}Device \\ Parameters\end{tabular}}} \\
                                 &                                                               &                                &                                                             &                                                                                                    &                                  &                               &                                                                                        \\ \bottomrule
\end{tabular}
    }

  % 3. 在最后统一添加标题和标签
  \caption{Construction framework of the PVK-KG and comparison with existing datasets.
  (a) The workflow for constructing the PVK-KG, illustrating the pipeline from data acquisition in academic repositories to information extraction. 
  We highlights how unstructured text is processed using a defined schema to generate structured triples (Entity-Relation-Source) focusing on fabrication and performance parameters.
  The following abbreviations are used in this figure: Fabrication (Fab.), Parameters (Para.), and Performance (Perf).
(b) Comparison of PVK-KG with other representative materials science knowledge graphs. 
We summarize key attributes including venue, year, scope, scale, and granularity, highlighting PVK-KG’s distinct automated approach and focus on perovskite device parameters compared to prior works.} 
  \label{fig:kg_supp2} % 这里的 label 用于引用整个组合
\end{figure*}

In Fig.~\ref{fig:kg_supp2}, we provide a comprehensive view of how the PVK-KG is constructed and how it positions itself relative to other established materials science datasets.
Fig.~\ref{fig:kg_supp2}(a) illustrates the end-to-end workflow for building the knowledge graph
The process begins with data acquisition from prestigious academic repositories and publishers, including Springer, arXiv, Nature, and Science.
These documents undergo a filtering process before being subjected to an automated extraction engine.
This engine, powered by large language models like GPT-4o, identifies and pulls specific fabrication and performance data directly from the text literature.
The Extraction Output demonstrates the model's ability to parse complex experimental details, such as: Fabrication Parameters: Identifying specific coating temperatures (e.g., $60^\circ\text{C}$) and freezing temperatures (e.g., $-20^\circ\text{C}$)
And then we organize information into a structured Entity-Relation-Source format.
For example, our pipeline identifies $SnO_2$ as the ETL Layer and links this fact directly to the source publication to ensure data provenance and knowledge grounding.
In Fig.~\ref{fig:kg_supp2} (b), we feature a comparative table that evaluates PVK-KG against other prominent materials science knowledge graphs like Propnet, Li-ion KG, and MatKG. 
This comparison highlights several advantages of the PVK-KG system.
Unlike many earlier databases that relied on manual curation or semi-automated methods, PVK-KG is fully automatic, utilizing LLMs for information extraction.
At the time of its $2025$ publication, PVK-KG contains $23,789$ entities and $22,272$ triples.
While most existing graphs focus on broad concepts, PVK-KG targets high-granularity Device Parameters. 
This focus is essential for the model to provide the precise, quantitative recipe adjustments needed for solar cell optimization.

\subsection{PVK-Sci Distribution}
\begin{figure*}
    \centering
    \includegraphics[width=1.0\linewidth]{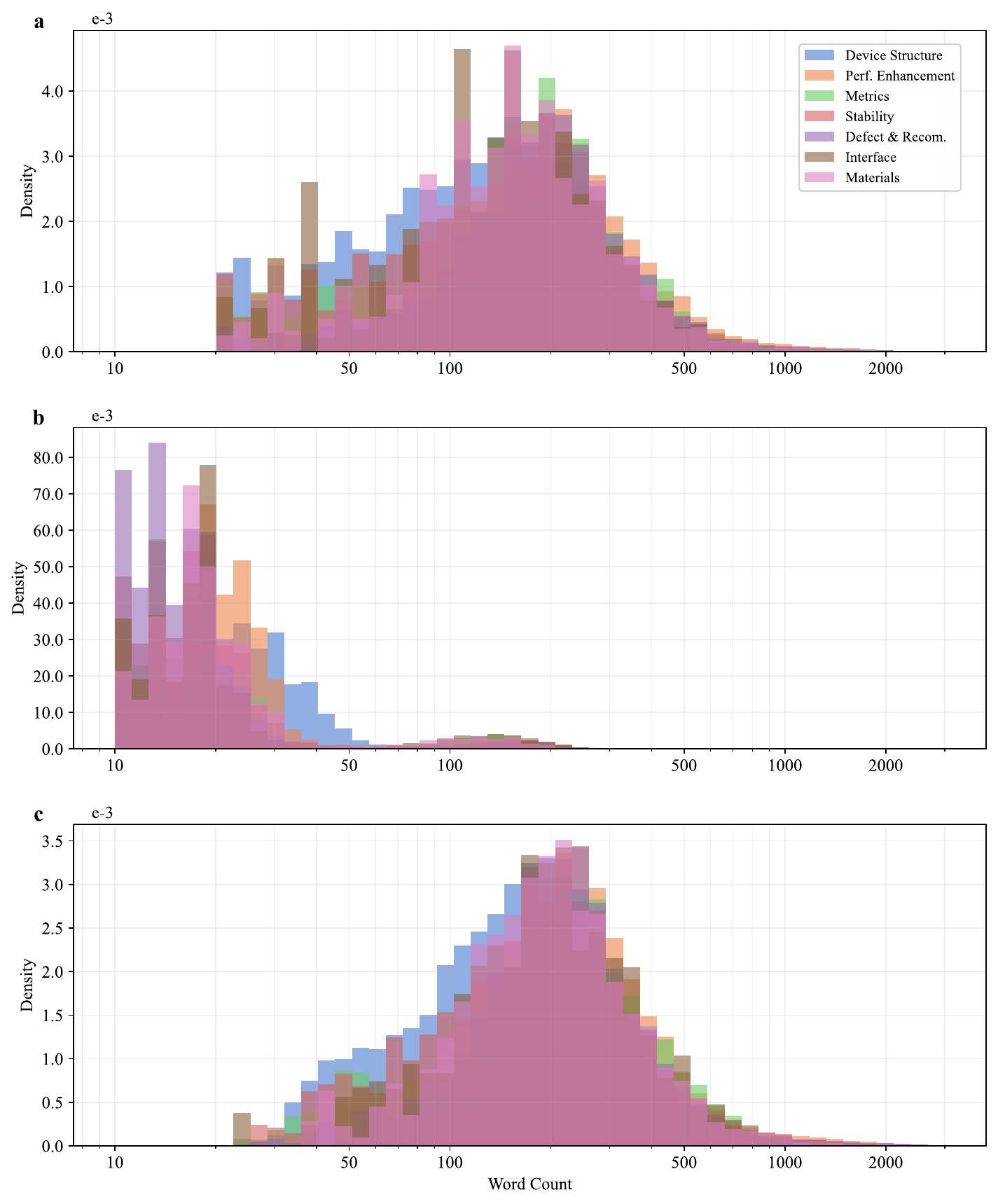}
    \caption{Length distribution analysis of PVK-Sci. 
    (a) Prompt length distribution across different categories. The data exhibits consistent patterns across categories, with sequence lengths predominantly falling between $100$ and $500$ tokens (median $\approx$ $400$). 
    (b) Response length distribution by category. 
    (c) Total sequence length distribution (prompt plus response). These statistics provide the basis for determining the maximum sequence length in our model design.}
    \label{fig:combined_length_distribution}
\end{figure*}

Fig.~\ref{fig:combined_length_distribution} presents a detailed statistical analysis of the PVK-Sci dataset, focusing on the token length distribution across various scientific categories.
Fig.~\ref{fig:combined_length_distribution} (a) displays the Prompt length distribution, showing the size of the initial input questions and scientific context for each research theme.
Fig.~\ref{fig:combined_length_distribution} (b) illustrates the Response length distribution, highlighting the depth and verbosity of the synthesized scientific answers provided by the model.
Fig.~\ref{fig:combined_length_distribution} (c) provides the Total sequence length distribution, which aggregates the prompt and response to show the complete data footprint of a single instruction-following pair.
The sequence lengths mainly fall within a range of $100$ to $500$ tokens. 
Furthermore, the analysis identifies that the median total sequence length is over $400$ tokens, suggesting a balanced complexity that is neither too brief for scientific reasoning nor excessively long for efficient training.

\subsection{Prompt Details}
In Figure~\ref{fig:prompt}, we present the details of the Prompt Engineering Framework, which serves as the operational core of the autonomous agents within the PVK-LLM system. 
In Figure~\ref{fig:prompt} (a), we utilize this prompt to extract the PVK-Sci, which is tasked with parsing paper content to answer standardized research questions. 
These prompts align with the seven thematic domains established in the earlier taxonomy, covering critical aspects such as device structures, performance metrics, defect passivation, and material features.
In Figure~\ref{fig:prompt} (b), Prompt B acts as an expert judge to validate the quality of the model's outputs. 
This prompt with LLMs compares generated responses against established ground truth across four dimensions: Accuracy, Completeness, Relevance, and Clarity. 
By assigning scores on a $1$–$5$ scale, this stage maintains the high standards required for academic-grade materials research.
In Figure~\ref{fig:prompt} (c), we show the prompt in the PVK-BO, which functions as a professional materials scientist to drive the autonomous cycle. 
It integrates observed experimental data and physical design principles with surrogate model predictions to select the most promising formulations. 
This agent specifically accounts for charge transport efficiency and model uncertainty, effectively balancing exploration and exploitation to pinpoint the global optima.

\begin{figure*}[p] % [p] 表示建议单独成页，因为三个框比较长
    \centering
    % --- 定义统一样式 ---
    \tcbset{
        mypromptstyle/.style={
            colback=gray!5!white,       % 浅灰背景
            colframe=gray!60!black,     % 深灰边框
            fonttitle=\bfseries\small\sffamily,
            listing only,
            listing options={
                basicstyle=\ttfamily\scriptsize, % 极小号字体，节省空间
                breaklines=true,
                columns=fullflexible,
                keepspaces=true,
                showstringspaces=false,
                aboveskip=1pt, belowskip=1pt
            },
            left=3pt, right=3pt, top=2pt, bottom=2pt,
            boxsep=1pt,
            arc=2pt, boxrule=0.5pt
        }
    }

    % --- 子图 1: 信息抽取 ---
    \begin{subfigure}[b]{\textwidth}
        \begin{tcolorbox}[mypromptstyle, title=\textbf{Prompt A: Information Extraction Agent}]
Answer the following questions based on the provided text.
{
  ``Device Structure": [``Q1: Summarize device config...", ``Q2: Precursor details...", ...],
  ``Performance": [``Q6: Improve VOC?", ``Q7: Improve FF?", ...],
  ``Defects": [``Q12: Passivation strategies?", ``Q13: Reduce recombination?", ...],
  ``Materials": [``Q17: HTL features?", ``Q18: ETL features?", ...]
}

Text: \{paper content\}

Response:
        \end{tcolorbox}
        \label{fig:prompt_extraction}
    \end{subfigure}
    
    \vspace{5pt} % 间距

    % --- 子图 2: 结果评估 ---
    \begin{subfigure}[b]{\textwidth}
        \begin{tcolorbox}[mypromptstyle, title=\textbf{Prompt B: Evaluation Agent}]
You are an expert evaluator. Your task is to compare a model's response to the ground truth answer and provide a detailed
evaluation.
Model's response:\{model response\} Ground truth:\{ground truth\}.

Please evaluate the model's response based on the following criteria:
1. Accuracy: How factually correct is the model's response compared to the ground truth?
2. Completeness: Does the model's response cover all the key points mentioned in the ground truth?
3. Relevance: How well does the model's response address the implied question or task?
4. Clarity: Is the model's response clear and easy to understand?

For each criterion, provide a score from 1 to 5, where 1 is the lowest and 5 is the highest. Also, provide a brief explanation
for each score.
Finally, give an overall score from 1 to 5 and a summary of your evaluation.
        \end{tcolorbox}
        \label{fig:prompt_evaluation}
    \end{subfigure}
    
    \vspace{5pt} % 间距

    % --- 子图 3: 材料优化 (新加的) ---
    \begin{subfigure}[b]{\textwidth}
        \begin{tcolorbox}[mypromptstyle, title=\textbf{Prompt C: Materials Optimization Agent}]
You are a professional materials scientist specializing in perovskite solar cells. 
Your goal is to select candidates to maximize \{target property\}.

\#\#\# Domain Knowledge

Key parameters: \{feature definitions\},
Ideal profiles: \{design principles\}.

\#\#\# Data Provided

1. Observed Data: \{observed data\},
2. Candidates: \{candidate data\},
3. Model Predictions: \{model predictions\}.

\#\#\# Task

Select top \{k\} formulations based on:
1. The PCE value,
2. Charge transport efficiency \& Recombination losses,
3. Model uncertainty (Exploration vs Exploitation).

Response Format:
Analysis: [Reasoning based on physics and data]
Selected Formulations: [List of IDs]
        \end{tcolorbox}
        \label{fig:prompt_optimization}
    \end{subfigure}

    \caption{The Prompt Engineering Framework. The pipeline consists of three stages: (a) Extracting structured knowledge from literature, (b) Validating the quality of extraction, and (c) Utilizing the knowledge and surrogate models to recommend optimal material formulations.}
    \label{fig:prompt}
\end{figure*}

\clearpage

\section{Device Fabrication in Wet-lab}
\label{sec:device_fabrication_in_wet_lab}
In the wet-lab experiments, we provide the recipe for the perovskite.
The specific details are provided below.

\noindent$\bullet$
\textbf{Substrate Cleaning and Hole Transport Layer (HTL) Deposition.} 
Pre-patterned fluorine-doped tin oxide (FTO) glasses are cleaned using ethanol for $20$ min in an ultrasonic bath, followed by plasma treatment for $15$ min. 
The hole transport layer (HTL) is fabricated using a self-assembled monolayer (SAM) solution of MeO-2PACz in a nitrogen-filled glovebox. 
The MeO-2PACz is dissolved in ethanol at a concentration of $0.5$ mg mL$^{-1}$. This solution was spin-coated onto the FTO substrates at $1500$ r.p.m. for $30$ s, followed by annealing at $100$ $^\circ$C for $10$ min.

\noindent$\bullet$
\textbf{Perovskite Absorber Layer Deposition.} 
A 1.67 M Cs$_{0.05}$MA$_{0.1}$FA$_{0.85}$PbI$_3$ perovskite precursor solution is prepared by dissolving $0.075$ mmol CsI, $0.15$ mmol MAI, $1.275$ mmol FAI, and $1.5$ mmol PbI$_2$ in $0.9$ mL of mixed solvent ($745$ $\mu$L DMF + $155$ $\mu$L DMSO). 
To improve film quality and provide adequate Pb$^{2+}$ sites for additive anchoring, $15$ mg MACl and $35$ mg PbI$_2$ are added to the solution. 
The precursor solution is stirred at $60$ $^\circ$C for $1$ h and filtered using a $0.22$ $\mu$m polytetrafluoroethylene (PTFE) membrane before use. 
The perovskite films are deposited via a two-step spin-coating process: $1000$ r.p.m. for $10$ s (acceleration: $500$ r.p.m./s) followed by $5000$ r.p.m. for $35$ s (acceleration: $1000$ r.p.m./s). 
During the second step, $150$ $\mu$L of anisole is dropped as an antisolvent $15$ s before the end of the program. 
The films are subsequently annealed at $120$ $^\circ$C for $10$ min.

\noindent$\bullet$
\textbf{Passivation Treatment (Epoch 0 vs. Epoch 1-3).} 
The passivation strategy is modified between the initial baseline (Epoch 0) and the optimization iterations (Epoch 1-3) to enhance interface quality and defect passivation.

For the baseline device (\textbf{Epoch 0}), a sequential passivation method is employed. 
A solution of $12$ mM 3MTPAI in $1$ mL isopropanol and a solution of $6$ mM PDAI2 in $1$ mL chlorobenzene are prepared separately. 
First, $60$ $\mu$L of the 3MTPAI solution is spin-coated onto the perovskite film at $4000$ r.p.m. for $25$ s, followed by annealing at $100$ $^\circ$C for $5$ min. 
Second, $60$ $\mu$L of the PDAI$_2$ solution is spin-coated at $4000$ r.p.m. for $25$ s, followed by annealing at $80$ $^\circ$C for $5$ min.

For the optimization rounds (\textbf{Epoch 1-3}), a mixed solution strategy is adopted. 
Four stock solutions are prepared in isopropanol (IPA): Solution A ($12$ mM 3MTPAI), Solution B ($6$ mM PDAI$_2$), Solution C ($4$ mM EDAI$_2$), and Solution D ($5$ mM PipDI).
Note that the solvent for PDAI$_2$ was changed to IPA for these iterations. 
These solutions are mixed according to specific ratios determined by the LLM. 
The resulting mixed passivation solution is spin-coated onto the perovskite films at $4000$ r.p.m. for $25$ s. The films were then annealed at $100$ $^\circ$C for 5 min in a nitrogen-filled glovebox.

\noindent$\bullet$
\textbf{Electron Transport Layer (ETL) and Electrode Deposition.} 
After cooling to room temperature, the substrates are transferred to a thermal evaporation system where a 30 nm C$_{60}$ film was deposited at a rate of $0.15$ \AA \  s$^{-1}$. 
Following C$_{60}$ deposition, the substrates are transferred to an atomic layer deposition (ALD) system (Picosun) to deposit a $25$ nm SnO$_2$ layer at $80$ $^\circ$C using tetrakis(dimethylamino) tin(IV) ($99.9999\%$) and deionized water as precursors.
Finally, a $120$ nm Cu electrode is deposited by thermal evaporation to complete the device.

\end{appendices}

\bigskip

\bibliographystyle{plainnat}
\bibliography{sn-bibliography}



\end{document}